\documentclass[USenglish,twocolumn]{article}

\ifx\directlua\undefined\ifx\XeTeXcharclass\undefined
  \usepackage[utf8]{inputenc}                           
  \else\RequirePackage[no-math]{fontspec}[2017/03/31]\fi 
  \else\RequirePackage[no-math]{fontspec}[2017/03/31]\fi 
\usepackage[sort&compress,square,numbers]{natbib}
\usepackage[big,online]{dgruyter}
\usepackage{caption}
\usepackage{graphicx}
\usepackage{xcolor}
\usepackage{soul}
\soulregister{\cite}7

\theoremstyle{dgthm}

\theoremstyle{dgdef}

\begin{document}

\articletype{Perspective}

\author[1,2]{Meng Tian}
\author[1,2]{Guanyu Han}
\author[1,2]{Ziyao Feng}
\author[1,2]{Yu Liu}
\author[1,2]{Yu Wang}
\author[1,2]{Wenjun Deng}
\author[3]{Ayed Al Sayem}
\author*[4]{Qiushi Guo} 
\affil[1]{Photonics Initiative, Advanced Science Research Center, City University of New York, NY, USA; Physics Program, Graduate Center, City University of New York, 365 5th Ave, New York, 10016, NY, USA}
\affil[3]{Nokia Bell Labs, NJ, United States}
\affil[4]{Photonics Initiative, Advanced Science Research Center, City University of New York, NY, USA}

\title{Lithium niobate quadratic integrated nonlinear photonics: enabling ultra-wide bandwidth and ultrafast photonic engines}
\runningtitle{}
\abstract{Integrated photonic coherent light sources capable of generating emission with broad spectral coverage and ultrashort pulse durations are critical for both fundamental science and emerging technologies. In this Perspective, we start by discussing emerging quantum and classical photonic applications from the standpoint of operating wavelength and timescale, highlighting the technological gaps that persist in current integrated photonic light sources. Next, we introduce the unique properties of lithium niobate-based integrated quadratic nonlinear photonics, and discuss several promising strategies that exploit this platform to realize wavelength-tunable continuous wave light sources and broadband, ultra-short light pulse generation. We also assessed their advantages and limitations while discussing potential solutions. Finally, we outline future prospects and challenges that need to be addressed, aiming at inspiring continued research and innovation in this rapidly evolving field.}
\journalname{Nanophotonics}

\journalyear{2025}
\journalvolume{aop}

\maketitle

\vspace*{-6pt}

\section{Introduction} 
\vspace*{-3pt}
As two fundamental properties of light, wavelength and timescale have been central to many advances in both science and technology. Historically, access to a broad spectral range allows for the investigation of emission and absorption features of diverse materials—crucial for spectroscopy and precision sensing\cite{theophile2012infrared}. It has also revolutionized modern communications through optical wavelength-division multiplexing (OWDM), supporting high-data-rate information transmission and processing \cite{ishio2003review,brackett2002dense,bergano2002wavelength}. Meanwhile, ultrashort-pulse lasers have made it possible to observe and measure natural phenomena with extreme temporal resolution \cite{fleming1986chemical,zewail2000femtochemistry,gao2014single}.
\vspace{1.5mm}

\begin{figure*}[ht]
\centering
\captionsetup{justification=centering}
\includegraphics[width=1\linewidth]{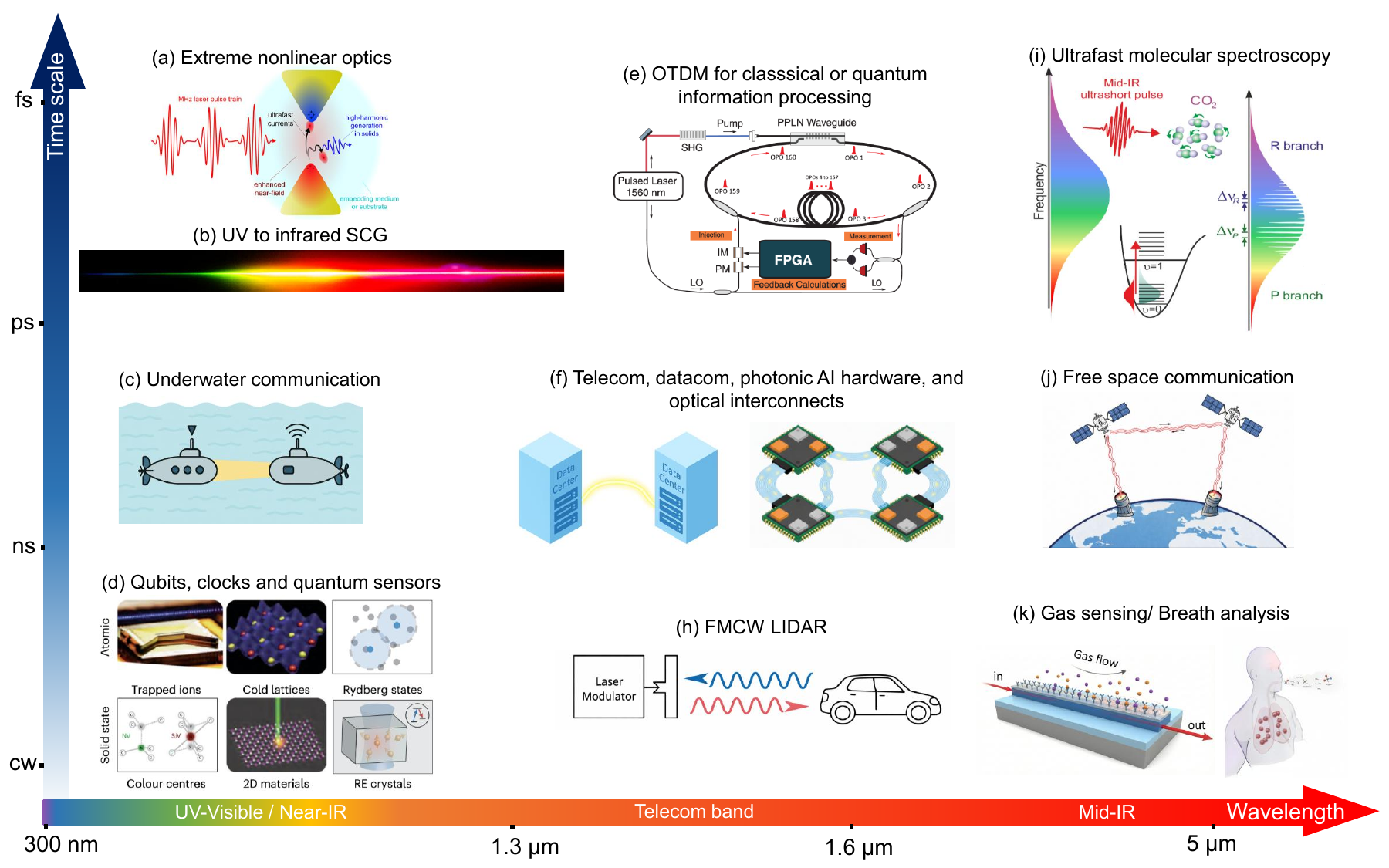}
\caption{Non-exhaustive summary of emerging classical and quantum photonic applications across different wavelengths and timescales. Sub-figure (a) is adapted from Ref. \cite{schotz2019perspective}; (b) is adapted from Ref. \cite{epping2015chip}; (d) is adapted from Ref. \cite{lu2024emerging}; (e) is adapted from Ref. \cite{mcmahon2016fully}; (i) is adapted from Ref. \cite{lanin2014time}. Subfigures (c), (f), (h), (j) and (k) were generated with the assistance of ChatGPT for illustrative purposes. It ought to be noted that many applications can span a large portion of the wavelength and timecale, either inherently or depending on the specific application instance.} \label{Fig1}
\end{figure*}

Today, we are transitioning into a new era of the information age—driven by intelligent algorithms, advanced classical and quantum sensors and processors, ever-expanding computing resources, along with unprecedented vast amounts of data. These developments call for a new class of chip-scale coherent light sources that offer broader spectral coverage, shorter pulse durations, higher coherence, and dynamic reconfigurability. Figure \ref{Fig1} presents a non-exhaustive overview of emerging classical and quantum applications of coherent light sources from the perspectives of wavelength and timescale. Along the wavelength axis, lasers in the telecom band (1.3–1.6 $\mu$m) have long served as the backbone of telecommunications and data communications. However, new applications in AI infrastructure—including in-package optical interconnects \cite{choi2025scalable,hossein200614}, photonic AI accelerators \cite{shen2017deep,feldmann2021parallel,ou2025hypermultiplexed,peserico2023integrated,davis2025rf}, and photonic edge-computing architectures \cite{sludds2022delocalized,wu2025micro}—are now driving the need for optical amplifiers or light sources with broader spectral coverage, extending beyond traditional telecom bands to support significantly more information channels. In these short-reach systems, coherence requirements can be relaxed, while direct integration with electro-optic modulators is highly desirable for high-speed on-chip data encoding and processing. Moreover, the capability for dense wavelength-division multiplexing (DWDM) in the telecom bands has important implications for photonic quantum information processing, enabling high-rate generation of entangled photon pairs and frequency-bin qubits \cite{pang2025versatile}, as well as the realization of multiphoton entangled states \cite{reimer2016generation} and vacuum-squeezed modes \cite{yang2021squeezed, jahanbozorgi2023generation}.
\vspace{1.2mm}
 
In the shorter wavelength range, visible and near-infrared (NIR) light sources are increasingly important for quantum computing, metrology, and sensing, as many atomic and solid-state quantum systems—including color centers, trapped ions, quantum dots, and neutral atoms—exhibit optical transitions between 300 and 1,000 nm \cite{duan2024visible,lu2024emerging,lu2019chip}. The ability to amplify or generate multiple visible wavelengths is crucial for enhancing quantum sensor readout \cite{yang2024titanium,schall2025laser}, pumping and manipulating atoms and ions \cite{lu2020chip}, stabilizing optical clocks, and interrogating atomic transitions \cite{blumenthal2024enabling,blumenthal2020photonic}. Moreover, efficient coherent frequency conversion between visible and telecom bands is essential for interconnecting quantum systems into large-scale quantum networks \cite{kimble2008quantum,guo2016chip,lu2019efficient}. Beyond quantum technologies, visible photonics also benefits underwater optical communication, owing to the lower water absorption in the visible range \cite{kaushal2016underwater,sun2020review}. In the longer wavelength regime, mid-infrared (mid-IR) coherent sources beyond 2 $\mu$m can be used to fingerprint molecular rotational and vibrational transitions, thus enabling breath analysis \cite{selvaraj2020advances}, gas sensing \cite{popa2019towards}, environmental monitoring \cite{ycas2018high}, and chemical hazard detection \cite{haas2016advances}. Moreover, coherent light within the 3–5 $\mu$m atmospheric transparency window is also critical for free-space optical communications and high-speed data links \cite{hao2017mid,su2025demonstration,didier2023interband,lee2026suspended}.
\vspace{1.2mm}

Now, we turn to the vertical axis—timescale. The temporal characteristics of coherent light sources, spanning from continuous-wave (CW) operation to ultrashort femtosecond (fs) pulses, define their monochromaticity, temporal resolution, and peak intensity, with each regime enabling distinct forms of light–matter interactions and applications. For example, CW or quasi-CW light sources provide a steady or quasi-steady optical output with high temporal stability and narrow linewidths, which are essential attributes for optical sensing, frequency-modulated continuous-wave (FMCW) LiDAR, laser pumping, atomic manipulation, and high-resolution laser spectroscopy. The specific performance requirements of CW sources vary across applications. FMCW LiDAR systems, for instance, can afford to sacrifice some level of coherence in favor of a fast chirp rate with high linearity in the frequency sweep, which directly impacts ranging accuracy and update speed. In contrast, high-resolution laser spectroscopy and precision atomic measurements demand exceptionally high frequency accuracy and long-term stability to resolve narrow spectral features and maintain phase coherence over extended timescales. At a shorter time-scale, low-peak-intensity light pulses with nanosecond (ns) to around 10 picosecond (ps) durations generated directly by optical modulations are widely employed as data bits in high-bandwidth optical communication, optical interconnects, and emerging photonic AI accelerators. 
\vspace{1.1mm}

Progressing into the picosecond to femtosecond regime, the ability to generate stable trains of femtosecond pulses enables the formation of optical frequency combs (OFCs), which have become indispensable tools for precision measurements \cite{cundiff2003colloquium,diddams1999broadband,gaeta2019photonic}. An OFC provides a phase-coherent set of equally spaced spectral lines that function as an “optical ruler” enabling precision measurements of unknown optical frequencies, high-resolution dual-comb molecular spectroscopy \cite{suh2016microresonator,picque2019frequency,dutt2018chip,yu2018silicon}, and astronomical applications such as exoplanet detection through ultra-precise Doppler spectroscopy \cite{suh2019searching}. For these applications, high temporal coherence of the pulse train is essential, as it directly determines the linewidth, stability, and spectral purity of the comb lines, and thus the achievable measurement accuracy. On the other hand, ultrashort optical pulses can be used for another powerful information multiplexing architecture known as optical time-division multiplexing (OTDM) \cite{tucker2002optical,hamilton2002100, willner2013all}, which has recently attracted significant attention in both classical and quantum information processing \cite{marandi2014network,li2024deep,mcmahon2016fully,inagaki2016coherent,yokoyama2013ultra,asavanant2019generation,kaneda2019high,larsen2019deterministic,erhard2020advances}. By encoding information into short temporal bins and densely packing them in the time domain, OTDM offers a pathway to dramatically increase the overall data rate, while easing the coupling of temporal bins using optical delay lines. Compared with OWDM, a distinctive feature of OTDM is its ability to exploit the high peak intensity of ultrashort light pulses to induce sufficient optical nonlinearities within the system. This unique capability allows one to harness optical nonlinearity in resonator-based OTDM architectures to perform computational operations \cite{marandi2014network,li2023all,choi2024photonic} or to efficiently generate non-classical states with a high data rate in a single-pass OTDM system \cite{asavanant2019generation,larsen2019deterministic,erhard2020advances,nehra2022few}. Furthermore, the interaction between high-peak-intensity picosecond or femtosecond pulses and nonlinear media can be leveraged to generate new optical frequencies that are otherwise inaccessible using conventional lasers. Such frequency synthesis can arise from a range of nonlinear processes, such as second-harmonic generation, sum- and difference-frequency generation, and supercontinuum generation (SCG) \cite{boyd2008nonlinear}.
\vspace{1.2mm}

At the extreme limit, the ability to synthesize few- or even single-cycle light pulses enables the direct observation and control of electron dynamics in atoms, molecules, and solids in real time \cite{krausz2009attosecond,goulielmakis2007attosecond,hui2022attosecond}. Moreover, their exceptionally high peak powers—several orders of magnitude greater than those of CW sources—usher in the regime of extreme nonlinear optics. In this regime, light-matter interactions become highly nonlinear and nonperturbative. The intense oscillating optical field governs the motion of bound electrons: they are tunnel-ionized, accelerated, and subsequently recombined with their parent ions when the field reverses, leading to phenomena such as high-harmonic generation (HHG)—the emission of a train of attosecond light bursts in the extreme ultraviolet (XUV) or even soft X-ray spectral regions \cite{paul2001observation,hentschel2001attosecond,drescher2001x}. 
\vspace{1.2mm}

In recent years, significant progress has been made in integrated photonic light sources, such as chip-scale CW or pulsed lasers \cite{coldren2012diode,corato2023widely,liu2024fully,yang2024titanium,wang2023photonic,xiang20233d,davenport2018integrated,zhang2020integrated}, optical amplifiers \cite{kuznetsov2025ultra,riemensberger2022photonic,zhao2023large,heydari2023degenerate,ye2021overcoming,liu2022photonic,bradley2011erbium,dean2026low,kellner2025low}, optical parametric oscillators (OPOs) \cite{lu2021ultralow,bruch2019chip,lu2020chip,pidgayko2023voltage}, and optical frequency combs based on nonlinear optical resonators \cite{gaeta2019photonic,bruch2021pockels} or electro-optic effects \cite{zhang2019broadband,rueda2019resonant}. These on-chip light sources offer key advantages in compactness, portability, and most importantly, significantly reduced pump power requirement compared to traditional tabletop free-space or fiber-based laser systems. Despite these advances, two major challenges still persist in terms of both the operating wavelength range and timescale. First, existing integrated photonic CW light sources generally lack the capability for fast, wide-range wavelength reconfiguration and switching. This limitation stems from two major challenges in conventional integrated photonic platforms: 1) the absence of broadband gain mechanisms for wide wavelength coverage; 2) the conventional electrical tuning approaches in integrated photonics—primarily electro-optic and thermo-optic effects—are too weak to induce sufficient perturbations in the photonic structure to achieve large wavelength shifts in on-chip light sources. 
\vspace{1.5mm}

Second, on the ultrafast timescale, there is still a lack of efficient on-chip pulsed light sources capable of delivering ultrashort pulses with high peak intensity—an essential requirement for directly driving other nonlinear processes on chip, such as on-chip supercontinuum generation, pulse compression, etc. For pulsed sources based on nonlinear resonators, this limitation arises from the instantaneous nature of parametric gain due to Kerr and quadratic ($\chi^{(2)}$) optical nonlinearity, which is not suitable for storing pump energy. This is in stark contrast to ultrafast mode-locked lasers, in which the signal can extract energy from the active laser gain medium over an extended interaction time due to the long excited-state lifetime of laser gain media.
\vspace{1.5mm}

Finally, integrated photonic light sources have conventionally been developed as standalone devices. However, many emerging applications demand system-level functionalities on chip, such as direct optical modulation of light sources for data encoding, frequency tuning and locking, new frequency generation, spectral broadening or pulse compression, and optical amplification. These evolving requirements highlight a pressing need for new material platforms that offer versatile, multi-functional capabilities to enable monolithically integrated photonic systems.

\begin{figure*}[ht]
\centering
\includegraphics[width=1\linewidth]{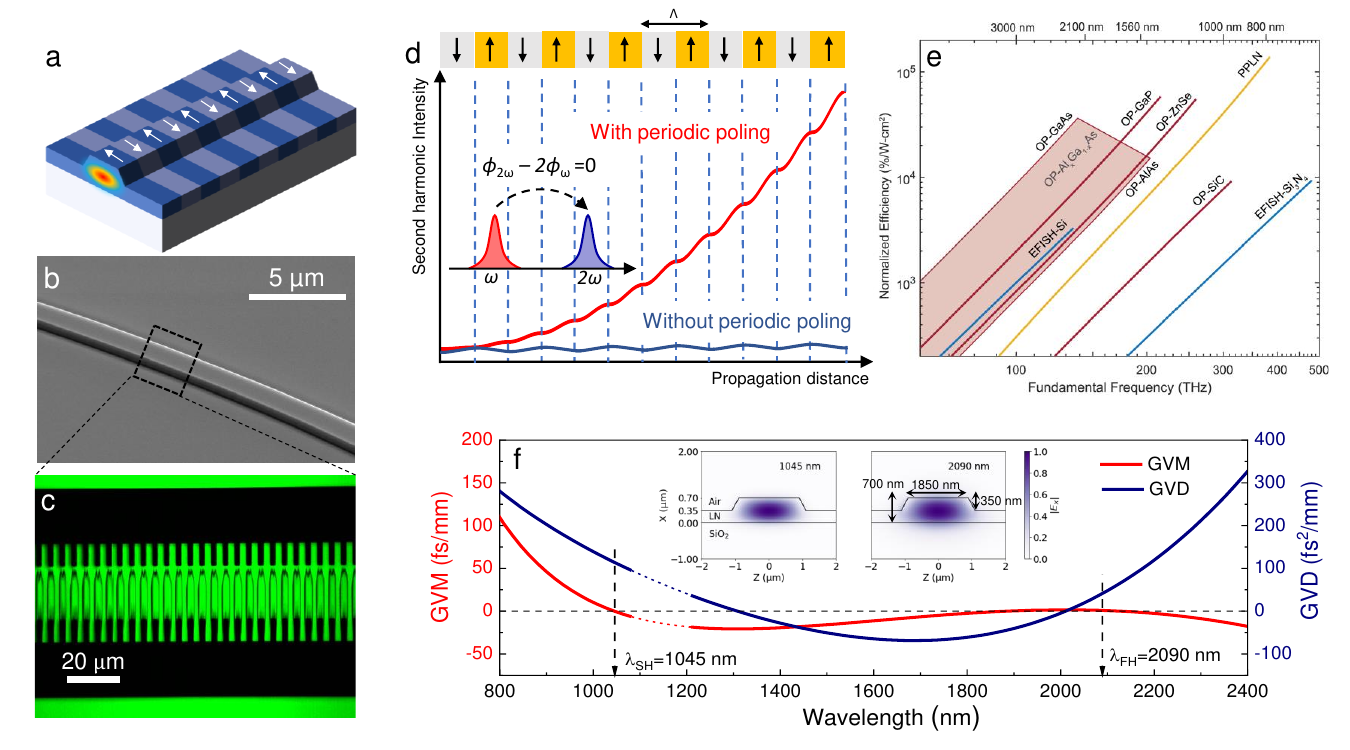}
\vspace{-5pt}
\caption{Lithium niobate $\chi^{(2)}$ integrated nonlinear photonics and its key features. (a) Schematic of PPLN nanophotonic waveguide. Arrows denote the orientations of ferroelectric domains.  (b) Scanning electron microscope (SEM) image of the PPLN nanophotonic waveguide. (c) Second harmonic microscope image of the periodic poling region in (b). (d) Simulated $\chi^{(2)}$ SHG process with (red) and without (blue) QPM (adapted from Ref. \cite{guo2022femtojoule}). (e) Calculated maximum normalized conversion efficiency as a function of FH frequency for different material platforms (adapted from Ref. \cite{jankowski2021dispersion}). (f) GVM (red) and GVD (blue) for the quasi-TE modes of a dispersion-engineered LN nanophotonic waveguide, with the corresponding cross-sectional geometry and optical mode profiles shown in the insets (adapted from Ref. \cite{guo2022femtojoule}).} \label{ppln}
\end{figure*}

\section{LN $\chi^{(2)}$ integrated nonlinear
photonics}
\vspace*{-4pt}

Lithium niobate (LN), a synthetic dielectric that exhibits both ferroelectric and piezoelectric properties, has long been recognized as a cornerstone material in optoelectronics due to its versatile functionalities \cite{boes2023lithium,chen2022advances}. For example, it exhibits a wide transparency window from 400 nm to 5 $\mu$m, strong electro-optic (E-O) effect (also termed as Pockels effect), large quadratic ($\chi^{(2)}$) optical nonlinear susceptibility, availability for quasi-phase-matching due to its ferroelectric nature \cite{fejer2002quasi}, as well as acousto-optic properties associated with its piezoelectricity \cite{weis1985lithium}. Moreover, it is compatible with rare-earth-ion doping (e.g., Er$^{3+}$, Yb$^{+}$), enabling its use as a solid-state laser gain medium or as a platform for quantum memories \cite{baumann1996erbium,brinkmann1994erbium}. The advent of wafer-scale thin-film lithium niobate (TFLN) \cite{levy1998fabrication,rabiei2004optical,zhang2017monolithic,poberaj2012lithium,zhu2021integrated} has transformed integrated photonics by combining LN’s strong $\chi^{(2)}$ nonlinearity and electro-optic effect within a compact nanophotonic platform, enabling device performance and functionalities that are difficult to achieve in conventional silicon or silicon nitride systems.
\vspace{1.5mm}

First, as illustrated in Fig. \ref{ppln}a, the nanoscale geometry of TFLN nanophotonic waveguides provides strong spatial confinement of interacting waves, such as the fundamental harmonic (FH) and second-harmonic (SH), thereby greatly enhancing their nonlinear interactions. As shown in Fig. \ref{ppln}a–c, periodic inverting TFLN’s ferroelectric domains (periodic poling) periodically modulates the sign of $\chi^{(2)}$ nonlinear coefficient along the waveguide, compensating for phase mismatch between the interacting waves. This technique, known as quasi-phase matching (QPM), allows the $\chi^{(2)}$ interaction to accumulate constructively over long propagation lengths without back-conversion, in stark contrast to the unpoled cases (Fig. \ref{ppln}d). Owing to the combination of strong spatial mode confinement and QPM, periodically poled lithium niobate (PPLN) nanophotonic waveguides exhibit exceptionally high normalized nonlinear frequency conversion efficiency $\eta_0>1,000 \%$/W-cm$^{2}$ \cite{wang2018ultrahigh,zhao2020shallow,jankowski2020ultrabroadband,rao2019actively,mckenna2022ultra}, far surpassing what is conventionally achieved in bulk LN crystals or waveguides. Interestingly, as shown in Fig. \ref{ppln}e, $\chi^{(2)}$ processes can be significantly more efficient at shorter wavelengths (the maximum $\eta_0$ scales with $1/\lambda^4$), owing to stronger modal overlap between interacting waves at shorter wavelengths \cite{jankowski2021dispersion}. For example, $\eta_0$ at an FH wavelength of 980 nm can reach $\sim$33000$\%$/W-cm$^{2}$ \cite{park2022high}, which is approximately 6× higher than at 1550 nm, and nearly 16× higher than at 2090 nm \cite{jankowski2020ultrabroadband}.  Such an ultra-high $\eta_0$ enables a wide range of $\chi^{(2)}$-based frequency conversion processes—including second-harmonic generation (SHG), sum-frequency generation (SFG), difference-frequency generation (DFG), and optical parametric amplification (OPA)—to operate with low power requirements and a compact device footprint. 
\vspace{1.5mm}

Second, compared with other phase-matching techniques such as modal phase matching \cite{liu2021aluminum}, QPM relaxes the stringent constraints on waveguide cross-sectional geometry. Thus, it provides far greater flexibility for dispersion engineering—an essential factor for ultra-broadband and ultrafast operation. In LN nanophotonic waveguides, strong optical confinement enables precise control of group-velocity dispersion (GVD) and group-velocity mismatch (GVM) through tailoring the waveguide-geometry, allowing access to distinct $\chi^{(2)}$ nonlinear interaction regimes. For example, under CW pump, a low GVD near the signal/idler frequencies allows for a wide parametric gain spectrum near degeneracy, and a wide frequency tuning range of optical parametric amplifiers (OPAs) and optical parametric oscillators (OPOs). For pulsed operation, simultaneously achieving low GVD and low GVM (as shown in Fig. \ref{ppln}f) minimizes temporal broadening and walk-off between the FH (2090 nm) and SH (1045 nm) ultrashort pulses, resulting in a ``quasi-static'' interaction regime \cite{jankowski2022quasi}. This regime enables numerous applications such as ultra-broadband SHG \cite{jankowski2020ultrabroadband}, octave-spanning supercontinuum generation due to $\chi^{(2)}$ interactions \cite{jankowski2023supercontinuum,zhou2025quadratic}, intense and high bandwidth optical parametric amplification \cite{ledezma2022intense,jankowski2022quasi}, femtosecond all-optical switching \cite{guo2022femtojoule}, and few-cycle squeezed-light generation \cite{nehra2022few}. Moreover, operating an OPO in a regime with large GVM and low GVD can induce significant temporal walk-off between the fundamental and second-harmonic pulses. In this case, the FH light pulse can extract energy more efficiently from—or even strongly deplete—the SH pulse. This leads to the formation of walk-off induced soliton \cite{roy2022temporal} or simultons \cite{jankowski2018temporal} for the efficient generation of high-peak-power optical pulses, which will be further elaborated in the following sections.
\vspace{1.5mm}

Lastly, the $\chi^{(2)}$ nonlinearity of LN also gives rise to the electro-optic (Pockels) effect, whereby a static or low-frequency electric field modulates the refractive index experienced by an optical wave. Compared with E-O modulators made of bulk LN, the strong optical confinement in nanophotonic TFLN waveguides allows for much smaller electrode spacing, thus dramatically enhancing the overlap between the optical mode and the applied DC or RF field \cite{wang2018integrated,zhu2021integrated,hu2025integrated,boes2023lithium}. This leads to highly efficient electro-optic modulation, typically characterized by an exceptionally low half-wave voltage–length product ($V_\pi L$) \cite{wang2018integrated,zhang2021integrated}. With optimized device geometries, $V_\pi L$ values as low as 0.64 V·cm at 1550 nm \cite{jin2021efficient} and 0.17 V·cm at 450 nm \cite{xue2023full} have been achieved. Such an efficient electro-optic response of TFLN can be exploited for laser stabilization and locking \cite{xue2025pockels}, fast wavelength tuning and switching \cite{xue2025pockels,snigirev2023ultrafast,hu2021chip}, broadband frequency-comb generation and optical spectrum broadening\cite{yu2022integrated,hu2025integrated}, pulse broadening or compression \cite{kolner1989temporal,kolner2002space}, phase-locking of laser modes for generating ultrashort pulses \cite{guo2023ultrafast,ling2024electrically}, and can be seamlessly integrated with QPM $\chi^{(2)}$ nonlinear components to realize reconfigurable on-chip photonic systems for various applications.

\begin{figure*}[ht]
\centering
\includegraphics[width=0.95\linewidth]{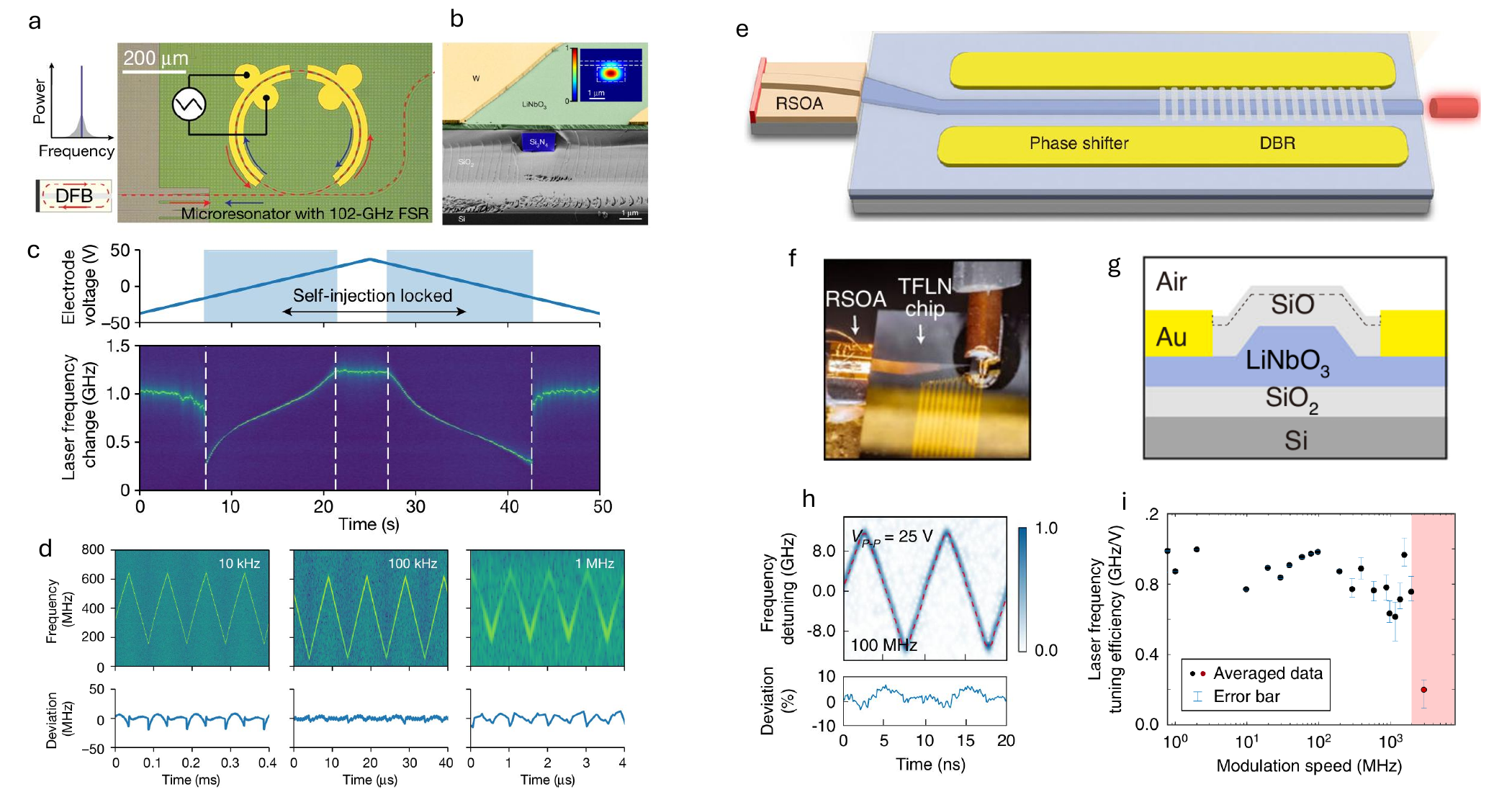}
\vspace{-5pt}
\caption{Fast E-O wavelength-tunable CW lasers on TFLN. (a) Schematic of fast wavelength-tunable self-injection laser. Laser wavelength tuning is achieved by applying a voltage signal on the tungsten electrodes (yellow) on TFLN. (b) False-colour SEM image of a heterogeneous Si$_3$N$_4$–LiNbO$_3$ waveguide cross-section. (c) Laser frequency change with time upon applying a linearly-modulated electrode voltage. The blue shaded region indicates the self-injection locking of the laser frequency to the microring resonance. (d) Time-frequency spectrograms of the heterodyne beatnotes at different modulation frequencies. The bottom shows the corresponding frequency deviation. (e) Schematic of a Pockels laser integrating a RSOA with external cavity DBR mirror. (f) Photo of the Pockels laser device. (g) The cross section of the e-DBR section. (h) Time-frequency spectrograms of the heterodyne beatnote and the corresponding frequency deviation. (i) Laser frequency tuning efficiency versus modulation speed. (a)-(d) are adapted from Ref. \cite{snigirev2023ultrafast}, (e)-(i) are adapted from Ref. \cite{xue2025pockels}.}  \label{tunablelaser}
\end{figure*}

\section{How LN $\chi^{(2)}$ integrated nonlinear
photonics fills the gap?  }
\vspace*{-4pt}
In this section, we outline several promising approaches that leverage the unique capabilities of LN integrated $\chi^{(2)}$ nonlinear photonics to achieve (1) widely wavelength-tunable CW light sources and (2) efficient generation of ultrashort pulses with broadband spectral coverage. We also assess the respective advantages and limitations of these sources. Section~3.1 introduces wavelength-tunable CW sources based on electro-optic effect, dispersion-engineered OPOs, and frequency-modulated OPOs. Section~3.2 focuses on broadband ultrafast light sources enabled by E-O modulation, mode-locking techniques, supercontinuum generation, two-color soliton pulse compression, and synchronously pumped OPOs operating in different regimes.

\subsection{Wavelength-tunable CW light sources }

\subsubsection{E-O wavelength-tunable CW lasers}
By hybrid integration of laser gain media with TFLN, the fast and efficient E-O response of LN nanophotonics can be leveraged to realize chip-scale, rapidly wavelength-tunable CW lasers. Wavelength-tunable narrow-linewidth CW lasers can be achieved either by E-O modulating the self-injection locking or modulating the laser external cavity. Figure \ref{tunablelaser}a shows a self-injection-locked laser architecture, in which an indium phosphide (InP) distributed feedback (DFB) laser is butt-coupled to a high-Q Si$_3$N$_4$ microring resonator with a bonded TFLN layer on top \cite{snigirev2023ultrafast}. The hybrid Si$_3$N$_4$–LN photonic platform (Fig. \ref{tunablelaser}b) combines the low-loss of Si$_3$N$_4$ waveguide and the efficient E-O tunability of TFLN. Back-scattering from the high-Q Si$_3$N$_4$ microresonator enables the self-injection locking of the DFB InP laser, yielding a laser linewidth of several kHz. Narrower laser linewidth can be achieved with higher Q-factor of the microresonator \cite{snigirev2023ultrafast,corato2023widely}. Applying an electric field across the LN region changes its refractive index via the linear E-O effect, thereby shifting the resonator’s resonant frequency and the laser’s output wavelength when operating within the self-injection locking range, as shown in Fig. \ref{tunablelaser}c. Because the Pockels effect in LN responds to the applied voltage nearly instantaneously, the laser frequency can be swept at very high rates (10$^{16}$ Hz/s) while retaining a linear, low-hysteresis response, as shown in Fig. \ref{tunablelaser}d. This stands in contrast to traditional tunable lasers where wavelength tuning is often dominated by thermal effects or piezoelectric actuators that are slower or exhibit mechanical instabilities. However, due to the narrow self-injection locking frequency range ($\sim$1 GHz), the linear frequency tuning range of the laser is restricted to hundreds of MHz \cite{snigirev2023ultrafast, li2023high}.
\vspace{1mm}

Compared with self-injection-locked lasers, external-cavity lasers offer wider tuning ranges, higher output power, and turnkey operation without the need for precise mode matching between separate cavities \cite{han2021electrically,li2022integrated,xue2025pockels,siddharth2025ultrafast}. Figure \ref{tunablelaser}e shows the structure of a Pockels laser \cite{xue2025pockels,li2022integrated}, which integrates a reflective semiconductor optical amplifier (RSOA) with an extended distributed Bragg reflector (eDBR) on the TFLN platform. Figure \ref{tunablelaser}f shows a photo of the entire laser device. Notably, instead of patterning the eDBR directly in the LN nanowaveguide, the grating is fabricated in the top SiO$_2$ cladding layer (cross section shown in Fig. \ref{tunablelaser}g). This approach produces a weak index perturbation and reduced scattering loss, resulting in an ultra-narrow reflection bandwidth and an intrinsic laser linewidth as low as 167 Hz \cite{xue2025pockels}. The eDBR grating can also be formed by etching small LN posts at both sides of the waveguide \cite{siddharth2025ultrafast}. The eDBR center frequency can be electrically tuned via the linear E-O effect in LN, while extended electrodes beyond the eDBR section enable simultaneous intra-cavity phase tuning. This ``co-tuning'' characteristic supports high tuning efficiency ($>$800 MHz/V) and extremely high tuning rate (10$^{19}$ Hz/s), as shown in Fig. \ref{tunablelaser}h and i. In addition, the mode-hop-free tuning range exceeds 10 GHz.
\vspace{1mm}

To enable a wider coarse wavelength-tuning range, a Vernier filter composed of two \cite{op2021iii,han2021electrically,li2022integrated} or three microrings \cite{ren2024widely} with slightly different radii can be used. An ultra-wide tuning range of 96 nm across the C, L and U bands has been demonstrated \cite{ren2024widely}. Besides, thermal-optic tuning can be combined with E-O tuning to allow both wide and ultrafast wavelength tuning \cite{li2022integrated}. Furthermore, a two-color tunable narrow-linewidth laser in the telecom and visible bands can be achieved by combining the E-O effect with nonlinear frequency conversion \cite{li2022integrated}.
\vspace{1mm}

The above-discussed lasers have profound impacts on a wide range of optical sensing and metrology applications. For example, the high tuning rate and linear response are particularly well suited for FMCW applications, such as coherent LiDAR ranging\cite{snigirev2023ultrafast,xue2023full}, where rapid and predictable frequency sweeps directly translate to high-resolution distance measurements. Moreover, high-speed intra-cavity phase modulation feature of the Pockels lasers allows for achieving Pound–Drever–Hall (PDH) laser frequency locking with dramatically simplified architecture. 
\vspace{1mm}

\begin{figure*}[ht]
\centering
\includegraphics[width=1\linewidth]{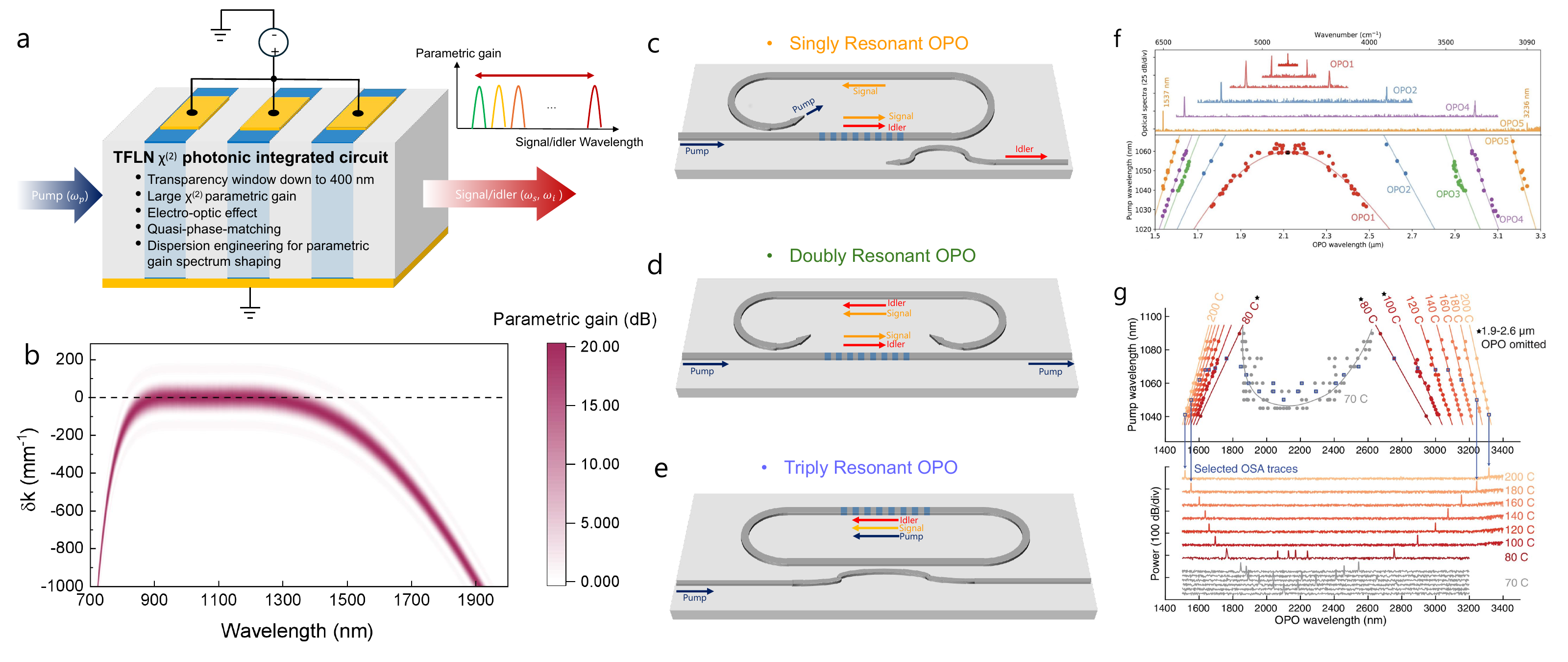}
\vspace{-20pt}
\caption{(a) Concept of the ultra-wide band light sources based on dispersion-engineered OPA or OPO. (b) Calculated optical parametric gain as a function of wavelength and externally induced perturbation wavevector $\delta k$ for a 7-mm-long PPLN nanophotonic waveguide with minimized $\beta_2$ and $\beta_4$, under a CW pump power of 1 W at 532 nm. (c–e) Examples of singly- (c), doubly-(d), and triply-resonant (e) TFLN nanophotonic OPOs. OPOs can also be implemented in linear cavity geometries using distributed Bragg reflectors \cite{kellner2025low}. (f) Upper panel: representative output spectra for a few doubly-resonant OPOs on the same chip. Lower panel: measured OPO wavelength (colored dots) versus the corresponding pump wavelength along with the theoretically calculated tuning curves (solid lines). (g) Upper panel: Singly-resonant OPO output wavelength tuning curves at different temperatures. Solid lines are theoretical tuning curves. Points marked with a blue outline have their output optical spectra plotted in the lower panel. (f) is adapted from Ref. \cite{ledezma2023octave}, (g) is adapted from Ref. \cite{hwang2023mid}.} \label{OPA}
\end{figure*}

\subsubsection{Widely wavelength-tunable CW OPOs}\label{section: CW OPO} 
Although the E-O effect enables fast laser wavelength tuning, the wavelength tuning range is still limited by the weak perturbation of refractive index of LN. Moreover, laser gain media based on semiconductors or doped-ions suffer from intrinsically limited gain bandwidths set by fixed energy levels, which fundamentally constrain the achievable wavelength tuning range. This limitation has hindered applications requiring full coverage of telecommunication bands or broadband laser spectroscopy of multiple ions, atoms, and chemical species.
\vspace{1.3 mm}

To overcome the abovementioned gain-bandwidth limitation, one can exploit the parametric gain mediated by virtual energy levels \cite{boyd2008nonlinear}, which enable energy transfer from a pump to signal and idler waves without being limited by material-defined real electronic transitions. In this context, a particularly promising strategy for generating broadband wavelength-tunable light is to leverage $\chi^{(2)}$ optical parametric gain in dispersion-engineered TFLN nanophotonic waveguides. Figure \ref{OPA}a illustrates this concept and highlights the unique advantages of TFLN nanophotonics for this particular application. Under optical pumping at frequency $\omega_\mathrm{p}$ (blue arrow shown in Fig.\ref{OPA}a), a QPM TFLN nanophotonic waveguide or resonator generates parametric gain at the signal ($\omega_\mathrm{s}$) and idler ($\omega_\mathrm{i}$) frequencies. The parametric gain coefficient is given by $g=\sqrt{\eta_0 P_\mathrm{pump}-(\Delta k/2)^2}$ \cite{ledezma2022intense}, in which $P_\mathrm{pump}$ is the pump power, $\eta_0$ is the normalized nonlinear frequency conversion efficiency, $\Delta k=k_\mathrm{p}-k_\mathrm{s}-k_\mathrm{i}-k_\mathrm{QPM}$ is the phase mismatch. Here, $k_\mathrm{j}=n_\mathrm{j}\omega_\mathrm{j}/c$ $(\mathrm{j}=\mathrm{p,s,i})$ are the wavevectors of the pump, signal, and idler waves, $n_\mathrm{j}$ are their effective refractive indices, and $k_\mathrm{QPM}=2\pi/\Lambda$ is the wavevector provided by QPM with period $\Lambda$.
\vspace{1.3mm}

Clearly, with fixed $P_\mathrm{pump}$, the peak parametric gain occurs at the phase-matching condition $\Delta k=0$. This condition defines a specific equilibrium dictated by the dispersion profile and the wavevector relationship of the interacting waves. Signal wavelength tuning can be achieved by perturbing one state to re-establish another one. Specifically, the modification of constituent wavevectors (such as the pump wavevector $k_\mathrm{p}$ or QPM wavevector $k_\mathrm{QPM}$) by external controls introduces a perturbation $\delta k$ to the original state. To re-establish $\Delta k=0$, the signal and idler frequencies $(\omega_\mathrm{s}$, $\omega_\mathrm{i})$ must shift to a new pair of solutions, while adhering to the energy conservation $\hbar\omega_\mathrm{p}=\hbar\omega_\mathrm{s}+\hbar\omega_\mathrm{i}$.
\vspace{1.3mm}

The tight spatial confinement for all interacting waves in TFLN nanophotonic waveguide enables precise dispersion engineering of their effective indices $n_{\mathrm{p}}$, $n_{\mathrm{s}}$, and $n_{\mathrm{i}}$. Compared to the case of bulk LN crystals or waveguides, this strong dispersion-engineering capability allows small variations in the phase mismatch $\Delta k$ to induce large shifts in the signal and idler frequencies $\omega_{\mathrm{s}}$ and $\omega_{\mathrm{i}}$, thereby enabling exceptionally wide wavelength tunability. In particular, near degeneracy ($\omega_{\mathrm{s}}\approx\omega_{\mathrm{i}}\approx\omega_{\mathrm{p}}/2$), the bandwidth of phase match $\Delta k\approx0$ depends predominantly on the group-velocity dispersion ($\beta_2$) and fourth-order dispersion ($\beta_4$)\cite{jankowski2021dispersion,han2026onchipelectricallyreconfigurableoctavebandwidth}. By judiciously designing the TFLN waveguide geometry and operating wavelength, $\beta_2$ and $\beta_4$ can be simultaneously minimized. This allows the condition $\Delta k\approx0$ to be preserved over an extremely broad wavelength range, yielding an ultra-wide degenerate parametric gain bandwidth from 900 nm to 1350 nm ($\sim$150 THz), as shown in Fig. \ref{OPA}b. Given such an ultra-broadband parametric gain around the degeneracy, a readily achievable $\delta k$ in principle leads to beyond-octave wavelength tuning of peak parametric gain wavelength from 720 to 1900 nm \cite{han2026onchipelectricallyreconfigurableoctavebandwidth}, shown in Fig. \ref{OPA}b. This exceptionally large wavelength tuning range covers key transition wavelengths of many visible photonic quantum systems and all telecom bands.
\vspace{1.3mm}

By combining the widely wavelength-tunable parametric gain with an integrated photonic resonator on the same TFLN chip, one can construct an OPO that serves as a broadly wavelength-tunable CW coherent light source. Figures~\ref{OPA}c–d illustrate several implementation examples of TFLN nanophotonic OPOs leveraging on-chip traveling wave resonators, which are discussed in detail in Ref.~\cite{lu2025photonic}. The singly-resonant OPO configuration (Fig.~\ref{OPA}c), in which only the signal wave (orange arrow) resonates in the cavity, offers the highest operational stability and the continuous wavelength tunability. This advantage arises from the absence of resonance constraints on the pump and idler waves, albeit at the cost of a higher oscillation threshold\cite{boyd2008nonlinear,suhara2003waveguide}. At the opposite extreme, the triply-resonant OPO (Fig.~\ref{OPA}e) \cite{lu2021ultralow,mckenna2022ultra}, in which the pump, signal, and idler waves are all resonant, enables the lowest oscillation threshold even down to 30 $\mu$W \cite{lu2021ultralow}. However, the strict resonance conditions can prevent smooth and continuous tuning of the pump and signal–idler wavelengths.
\vspace{1.3mm}

\textbf{OPO wavelength-tuning methods:} A practical wavelength tuning mechanism for TFLN nanophotonic OPOs is by varying the pump frequency $\omega_\mathrm{p}$, which introduces $\delta k=k_\mathrm{p}(\omega_\mathrm{p})-k_\mathrm{p}(\omega_\mathrm{p}\pm\Delta\omega)$ and results in a shift for the phase-matched solution ($\omega_\mathrm{s},\,\omega_\mathrm{i}$). With appropriate dispersion engineering, a modest change in $\omega_\mathrm{p}$ can translate into a large shift of $\omega_\mathrm{s}$ and $\omega_\mathrm{i}$. As shown in Fig. \ref{OPA}f, it was recently demonstrated that in a TFLN doubly-resonant OPO with a dispersion-engineered PPLN section, near the degeneracy, tuning the pump wavelength by only 30 nm around 1060 nm enabled a remarkably broad OPO signal and idler wavelength tuning range from 1.76 $\mu$m to 2.51 $\mu$m—covering over 750 nm \cite{ledezma2023octave}. Furthermore, in this work, ultra-broadband output tuning from 1.53 $\mu$m to 3.25 $\mu$m was achieved across five neighboring OPOs on a single nanophotonic chip, each with a small poling-period interval of 10 nm. Remarkably, the doubly-resonant OPO demonstrated in this work exhibits a low oscillation threshold of approximately 30 mW near degeneracy. This enables direct pumping with a CW DBR laser diode obviating the need for expensive high-power tunable laser sources \cite{ledezma2023octave,englebert2025topological}. Additionally, in an integrated LN singly-resonant OPO, by varying both the pump wavelength and the device temperature, the output of integrated LN OPO can be very widely tuned from 1.55 $\mu$m to 3.3 $\mu$m \cite{hwang2023mid}. Such a broad wavelength range in the near- to mid-IR spectral range allows the integrated LN OPO to be directly used for on-chip spectroscopic sensing of multiple chemical species. 
\vspace{1.5mm}

It should be noted that when tuning the pump frequency  $\omega_\mathrm{p}$, the signal frequency tuning slope  $\partial\omega_\mathrm{s}/\partial\omega_\mathrm{p}$ is given by the ratio of group-velocity differences $(1/v_\mathrm{i}-1/v_\mathrm{p})/(1/v_\mathrm{i}-1/v_\mathrm{s})$, while the parametric gain bandwidth scales inversely with $|(1/v_\mathrm{s}-1/v_\mathrm{i})|$ \cite{ledezma2023octave}. Thus, the design of wavelength-tunable OPOs requires maximizing the tuning slope, while simultaneously minimizing the risk of mode-hopping that arises from an excessively large parametric gain bandwidth. In doubly-resonant OPOs, additional design considerations are required to ensure that the resonance conditions for both signal and idler waves are met throughout the tuning range \cite{eckardt1991optical}. This involves optimizing the coupling between the pump, signal, and idler modes within the cavity, often by adjusting the cavity length and the resonant frequencies of the modes. Furthermore, advanced high-precision on-chip temperature control and active tuning  \cite{dacha2025frequency} can play a critical role in stabilizing the OPO, maintaining resonance alignment during wavelength tuning, and enlarging the wavelength tuning range combined with pump frequency tuning.
\vspace{1.5mm}

However, the pump frequency tuning typically necessitates widely tunable, mode-hop-free pump laser sources, which are often expensive and technically demanding. In this context, the TFLN platform offers a more integrated and agile alternative. By leveraging its efficient E-O effect, the phase-matching condition can also be efficiently modulated on chip, similar to the demonstrations in bulk QPM LN waveguides \cite{gross2002wide,o1999electro}. As illustrated in Fig. \ref{OPA}a, within each QPM period, the applied voltage selectively modifies the effective refractive indices of pump, signal and idler waves in one ferroelectric domain, whereas the refractive indices in the adjacent domain remain unchanged. This asymmetry prevents the mutual cancellation of index changes between adjacent domains, giving rise to a net wavevector perturbation $\delta k$ and thereby producing new solutions ($\omega_\mathrm{s},\,\omega_\mathrm{i}$) for the phase-matching condition. Note that the required $\delta k$ values shown in Fig. \ref{OPA}b can be achieved with practical on-chip applied voltages, owing to the efficient E-O effect. This underscores the promising prospect of achieiving electrically wavelength-tunable OPOs by combining dispersion engineering and E-O modulation. Because the E–O response is nearly instantaneous, this wavelength-tuning mechanism is particularly promising for realizing ultrafast wavelength-switched networks in data centers \cite{raja2021ultrafast}, FMCW LiDAR \cite{bianconi2025requirements}, and rapidly tunable light sources for high-speed spectroscopic sampling.
\vspace{1.5mm}

\textbf{Toward single-mode wavelength-tunable OPO:} Although dispersion-engineered integrated LN OPOs offer low pump thresholds and unprecedented wavelength tunability, the oscillation can become highly multi-mode with reduced temporal coherence when the OPO is pumped well above threshold. This effect is particularly pronounced near degeneracy, where the parametric gain bandwidth is large \cite{ledezma2023octave,hwang2023mid}. To address this issue, one approach is to introduce a wavelength-tunable auxiliary resonator as a narrow-band spectral filter to enforce single-mode oscillation \cite{tomazio2024tunable}. Alternatively, single-mode operation can be potentially achieved by significantly narrowing the parametric gain bandwidth—for instance, by maximizing the group-velocity difference between the signal and idler waves through careful dispersion engineering. Another effective strategy is to phase-couple the multiple oscillating longitudinal modes within the parametric gain bandwidth using intra-cavity E–O modulation \cite{diddams1999broadband}. When the driving frequency of the E-O phase modulator is slightly detuned from the OPO cavity free spectra range (FSR), a super-mode composed of one single longitudinal mode and its multiple sidebands can oscillate inside the cavity \cite{harris2003theory}. By demodulating the super-mode by applying an external modulator with opposite phase and appropriate modulation depth, a single longitudinal-mode oscillation can be established \cite{massey1965generation,osterink1967single}. 
\vspace{1.5mm}

\textbf{Toward even wider wavelength tunability:} Dispersion-engineered OPOs on TFLN can be readily scaled into more complex nonlinear optical systems to further extend the wavelength tuning range and spectral coverage. For example, by incorporating an additional PPLN section for broadband DFG within the OPO cavity \cite{stegeman2012nonlinear,koyaz2024ultrabroadband,li2025integrated}, the OPO can generate even longer-wavelength radiation, reaching the 4–5 $\mu$m regime relevant for molecular spectroscopy. Moreover, by cascading two dispersion-engineered, low-threshold OPOs on the same TFLN chip, the widely wavelength-tunable signal or idler from the first OPO can be used as the pump for the second OPO, thereby enabling an extremely broad wavelength tuning range on a single chip.

\subsubsection{E-O frequency-modulated OPOs}

\begin{figure}[ht]
\centering
\includegraphics[width=1\linewidth]{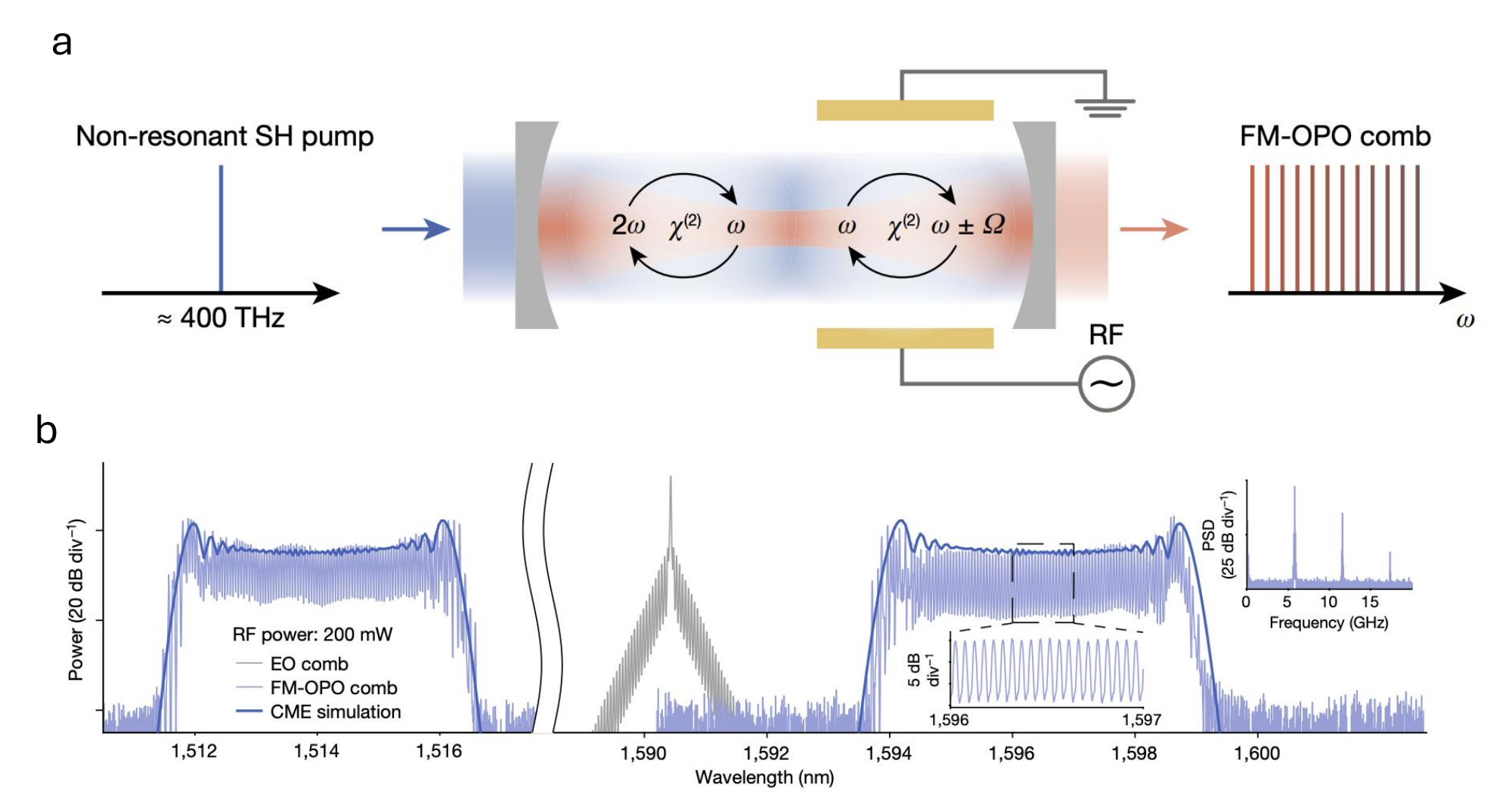}
\vspace{-20pt}
\caption{(a) Schematic illustration of the operating principle of the FM-OPO. (b) Output spectrum (blue) of the FM-OPO showing operation at the signal and idler frequencies. (a) and (b) are adapted from Ref. \cite{stokowski2024integrated}.}
\label{FM-OPO}
\end{figure}

Introducing intra-cavity phase modulation in an OPO can substantially broaden the OPO spectrum, generating a pair of wideband frequency combs centered at the signal and idler frequencies, as shown in Fig. \ref{FM-OPO} \cite{stokowski2024integrated} a and b. In the absence of parametric gain, the amplitudes of sidebands produced by electro-optic phase modulation inside a linear resonator typically decay exponentially with increasing frequency offset from the carrier \cite{zhang2019broadband}. In contrast, parametric gain in an OPO can coherent amplify these modulation sidebands. This results in remarkably flat and broadband spectra \cite{diddams1999broadband} around the signal and idler frequencies. For such a frequency-modulated (FM) OPO, the achievable bandwidth is ultimately limited by dispersion rather than resonator loss. In an FM OPO implemented on TFLN by Stokowski et al., a nearly flat-top spectral distribution spanning approximately 200 comb lines was demonstrated, along with a measured internal pump-to-comb conversion efficiency exceeding 93$\%$. Notably, the time-domain output of the FM-OPO exhibits a constant optical intensity without pulse formation, closely resembling that of FMCW lasers. This behavior originates from the dispersion-induced mismatch between the FSRs of the widely separated signal and idler waves, which enforces a persistent RF frequency detuning when the OPO operates in the non-degenerate regime. Compared to conventional FMCW lasers, however, FM-OPOs offer substantially greater spectral flexibility, enabled by spectrally tunable parametric gain as described in Section 3.1.2.

\subsection{Ultrashort pulsed light sources }\label{section: Ultrashort pulsed light sources}

\subsubsection{E-O frequency combs}\label{section: E-O comb}
A straightforward approach to generating short optical pulses on TFLN is to exploit its E-O property. In the simplest picture, when a sinusoidal microwave signal is applied to an E-O phase modulator, the induced optical phase shift varies periodically with the microwave amplitude. This leads to the generation of new optical frequency components (sidebands) spaced by the modulation frequency. Owing to its efficient electro-optic response and flexibility in dispersion engineering, TFLN offers remarkable potential for generating very broadband E–O frequency combs and ultrashort optical pulses with a very compact device footprint.
\vspace{1.5mm}

\begin{figure}[ht]
\centering
\includegraphics[width=1\linewidth]{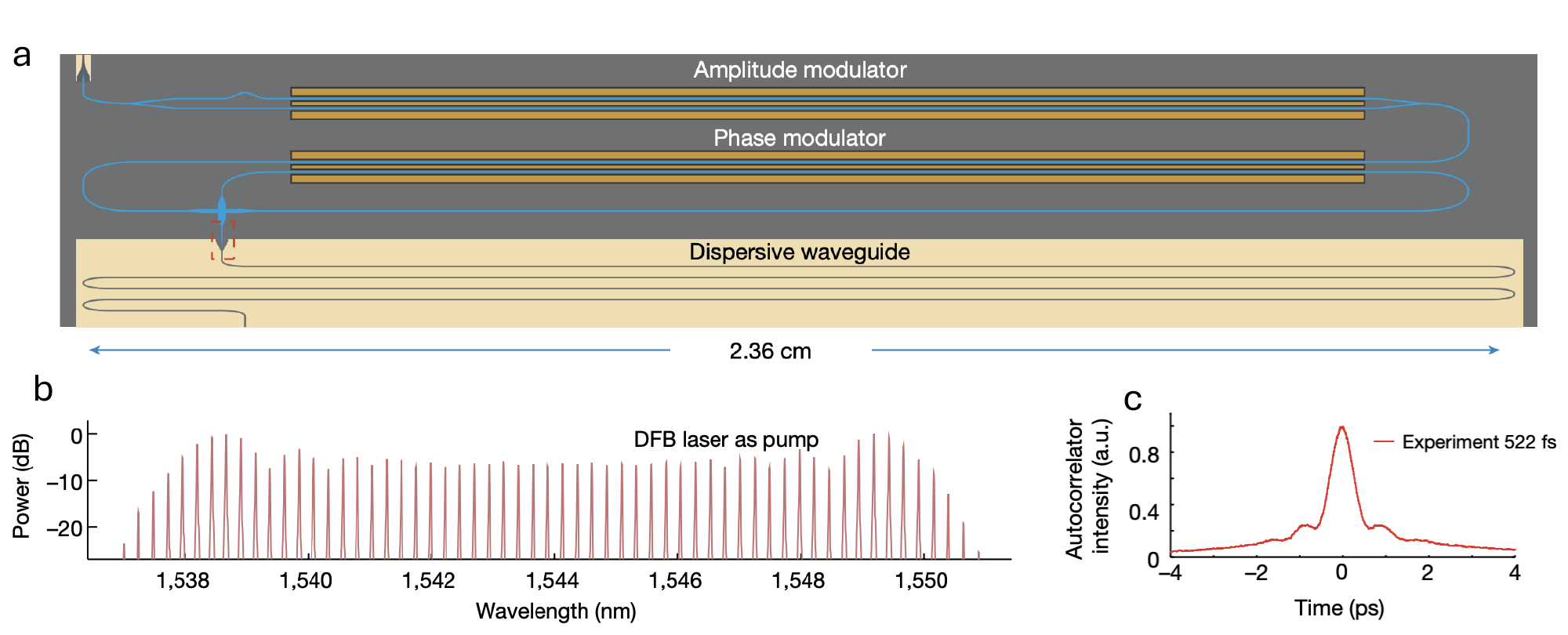}
\vspace{-20pt}
\caption{Time-lens-based E-O comb generator on TFLN. (a) Optical microscope image. (b)  Optical spectrum of the time-lens-based E-O comb generator. (c) Autocorrelation traces of the femtosecond optical pulse. (a), (b) and (c) are adapted from Ref. \cite{yu2022integrated}.} \label{timelens}
\end{figure}

An obstacle to achieving ultrashort optical pulse generation using the E-O modulation lies in the large linear frequency chirp introduced by phase modulation. To overcome this limitation, a waveguide-based time-lens system can be implemented on TFLN to temporally compress the chirped pulse even into the femtosecond regime \cite{mengjie2022timelens}, as schematically shown in Fig. \ref{timelens}a. This configuration operates in direct analogy to spatial optics: an amplitude modulator defines a temporal aperture, the phase modulator acts as the lens in the temporal domain, and a dispersive waveguide or grating provides the required focal distance. Together, these elements focus the optical pulse in time, enabling the experimental generation of an ultrashort optical pulse train with a 10-dB bandwidth of 12.6 nm (Fig. \ref{timelens}b) and pulse width as short as 522 fs (Fig.\ref{timelens}c) \cite{mengjie2022timelens}. The achievable pulse width of E–O time-lens systems can be further reduced by increasing the modulation depth through optimized waveguide or phase modulator design \cite{wang2025highly,lee2025ultra,zhang2023power}, employing higher RF modulation frequencies, or pumping nonlinear optical effects on TFLN for further pulse compression \cite{cheng2024frequency,ren2025few}. Such short pulses can be used as a pump for generating pure photons using non-linear parametric photon pair sources \cite{xin2022spectrally}, offering monolithic integration of pump and photon-pair source on LN.
 \vspace{1.5mm}
 
\begin{figure}[ht]
\centering
\includegraphics[width=1.01\linewidth]{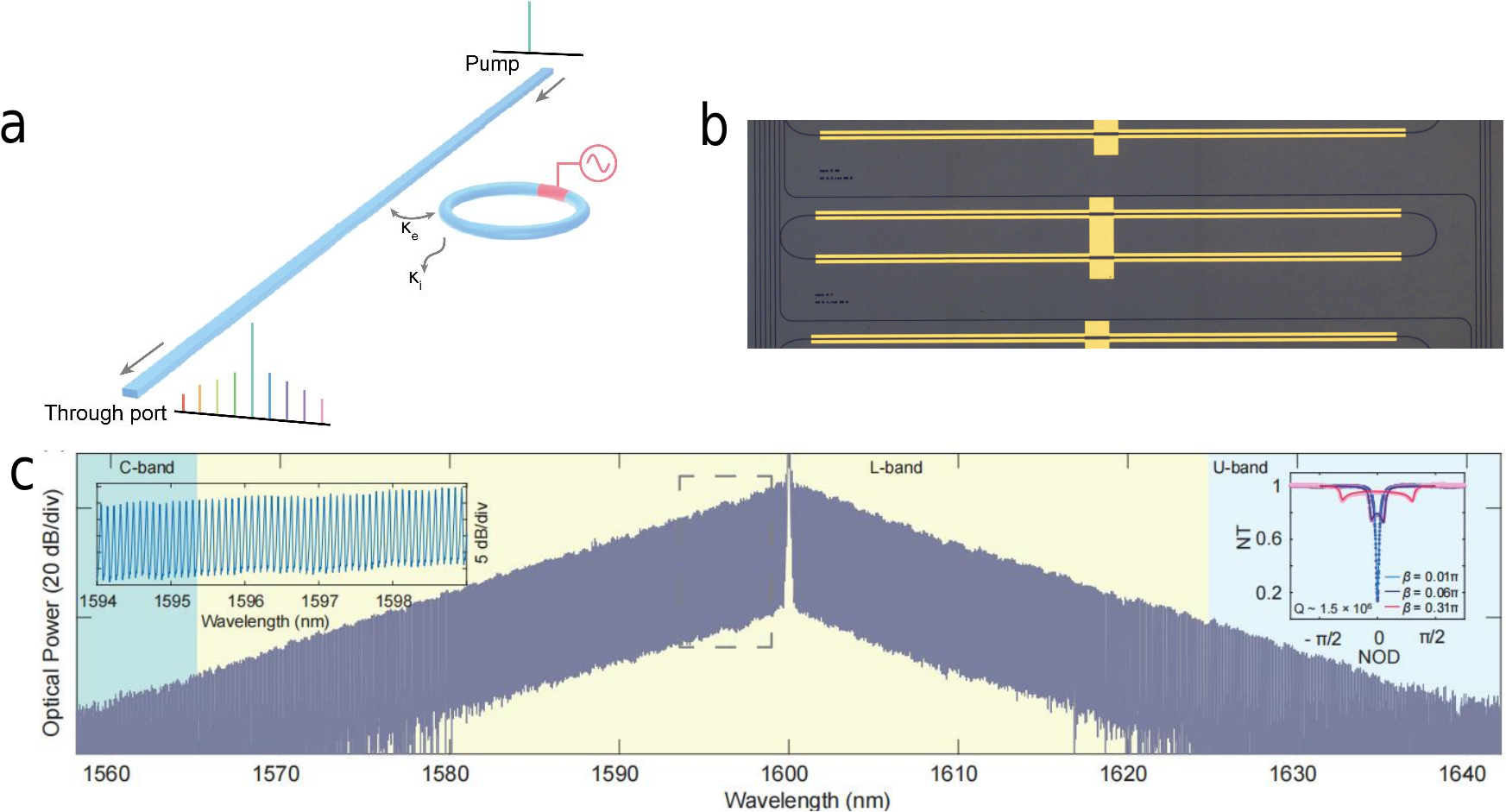}
\vspace{-20pt}
\caption{Single-resonator E-O comb generator on TFLN. (a) Schematic and (b) optical microscope image of single-resonator-based E-O comb generator. (c) Measured optical spectrum of the generated E-O comb. (a) is adapted from Ref. \cite{hu2022high}, (b) and (c) are adapted from Ref. \cite{zhang2019broadband}. }  \label{EOcombSR}
\end{figure}

Compared to a single-pass waveguide-based E-O time lens, resonator-based E-O combs can offer a much broader optical bandwidth. By placing an E-O modulator inside an optical resonator (Fig. \ref{EOcombSR}a), light circulates and passes through the modulator multiple times. This cascading modulation allows the comb to expand dramatically while simultaneously reducing phase noise, as the resonator promotes coherent buildup of sidebands and suppresses uncorrelated phase fluctuations originating from the CW pump. In this case, the dominant noise contribution arises from the RF modulation itself, whose fluctuations are multiplicatively transferred through the cascading modulation \cite{zhuang2023electro}. A notable example was demonstrated by Zhang et al. in 2019 \cite{mianzhang2019singlecavitycomb}, who utilized a single resonator in a TFLN nanophotonic platform (Fig. \ref{EOcombSR}b) to generate an E-O comb spanning 80 nm, as shown in Fig. \ref{EOcombSR}c. Broader comb spans and reduced microwave drive power can be achieved using a triply-resonant configuration incorporating an on-chip microwave resonator, in which two optical modes and one microwave mode are simultaneously resonant. This approach results in dramatically enhanced E-O interaction. Using this approach, a recent work demonstrated a 450-nm-wide comb with over 2,000 lines on a thin-film lithium tantalate platform \cite{zhang2025ultrabroadband}. With optimized dispersion engineering, the bandwidth can be further extended toward a full octave.

 \vspace{1.5mm}

\begin{figure}[ht]
\centering
\includegraphics[width=1\linewidth]{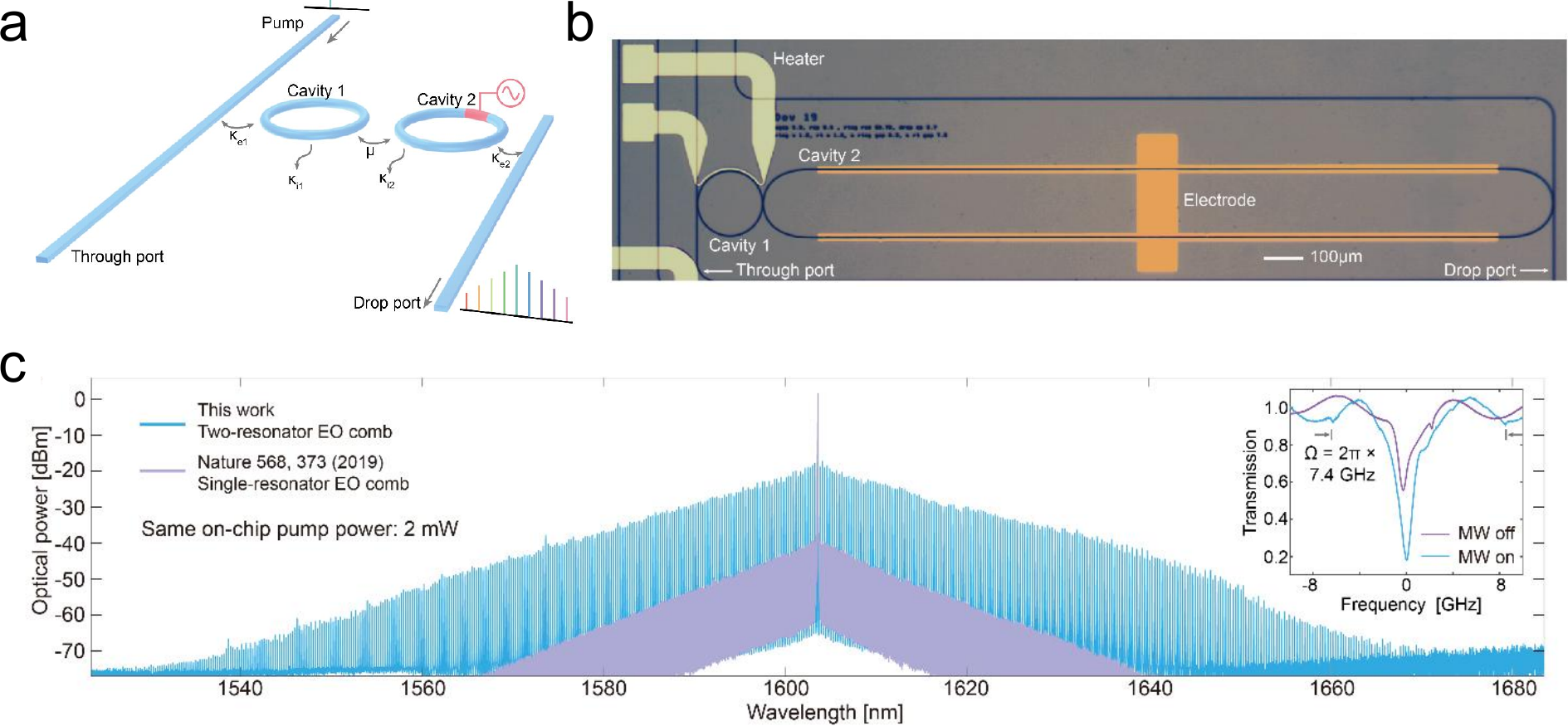}
\vspace{-10pt}
\caption{Dual-resonator E-O comb generator on TFLN. (a) Schematic and (b) optical micrograph of dual-resonator-based E-O comb generator. (c) Measured output comb spectra of single- (purple) and dual-resonator (blue) E-O comb generator. (a) (b) and (c) are adapted from Ref. \cite{hu2022high}.}  \label{EOcombDR}
\end{figure}
\vspace{1.5mm}

However, a single-resonator E-O combs often yields low pump-to-comb conversion efficiency ($\sim$0.3\%), which stems from a severely under-coupled cavity resonance induced by the strong microwave drive signal \cite{hu2022high}. In other words, most of the pump energy leaked out of the cavity without generating comb lines. An effective solution is to introduce an additional auxiliary resonator \cite{yaowen2022dualcavitycomb,buscaino2020design} to enforce pump energy storage. As shown in Fig. \ref{EOcombDR}a and b, cavity 1, which serves as the auxiliary resonator, is designed to over-couple only the pump resonance of the main racetrack cavity (cavity 2), thereby rejecting the remaining spectral modes required for frequency comb generation. When a strong microwave signal is applied to the main cavity, these two resonators form a critically coupled system and therefore the pump energy can be efficiently stored inside the dual-resonator structure, which significantly boosts comb generation efficiency. This approach has led to a 100-fold improvement in conversion efficiency (from 0.3$\%$ to 30$\%$) and a 2.2-fold increase in bandwidth (up to 132 nm) compared to single-resonator comb generators (Fig. \ref{EOcombDR}c). Such a high-efficiency E-O comb gives rise to femtosecond pulse (336 fs) generation on chip, with a high intra-cavity pulse peak power of 85 W \cite{hu2022high}. 
\vspace{1.5mm}

Recent investigations into cavity electro-optic dynamics have established a general framework for strong-coupling, high-bandwidth E–O modulation \cite{lei2025strong}, offering deeper insights into E–O comb generation and coupling mechanisms. This framework reveals a direct link between higher-order E–O comb dynamics and synthetic frequency-dimension band structures, and extends to the high-bandwidth regime where arbitrary microwave drive waveforms can be applied. Using a machine-learning (ML)–based inverse design process, non-flatness in cavity E–O combs was corrected: given a target comb shape, the ML model outputs the optimized high-bandwidth microwave waveform required to generate it. Combining this ML-optimized drive with detuning-induced frequency boundaries produced a flat-top comb with a tenfold improvement in flatness (0.03 dB/line over 200 lines). The approach also uncovered new physical insights—linking the drive pulse’s temporal edges to comb spectral slope—enabling arbitrary, programmable shapes such as “single-side-flat” or “unequal-arm-flat” combs. These advances mark a new frontier in cavity electro-optic modulation, centered on ML-programmable frequency combs for topological photonics and quantum photonic computing.

\begin{figure*}[ht]
\centering
\includegraphics[width=0.8\linewidth]{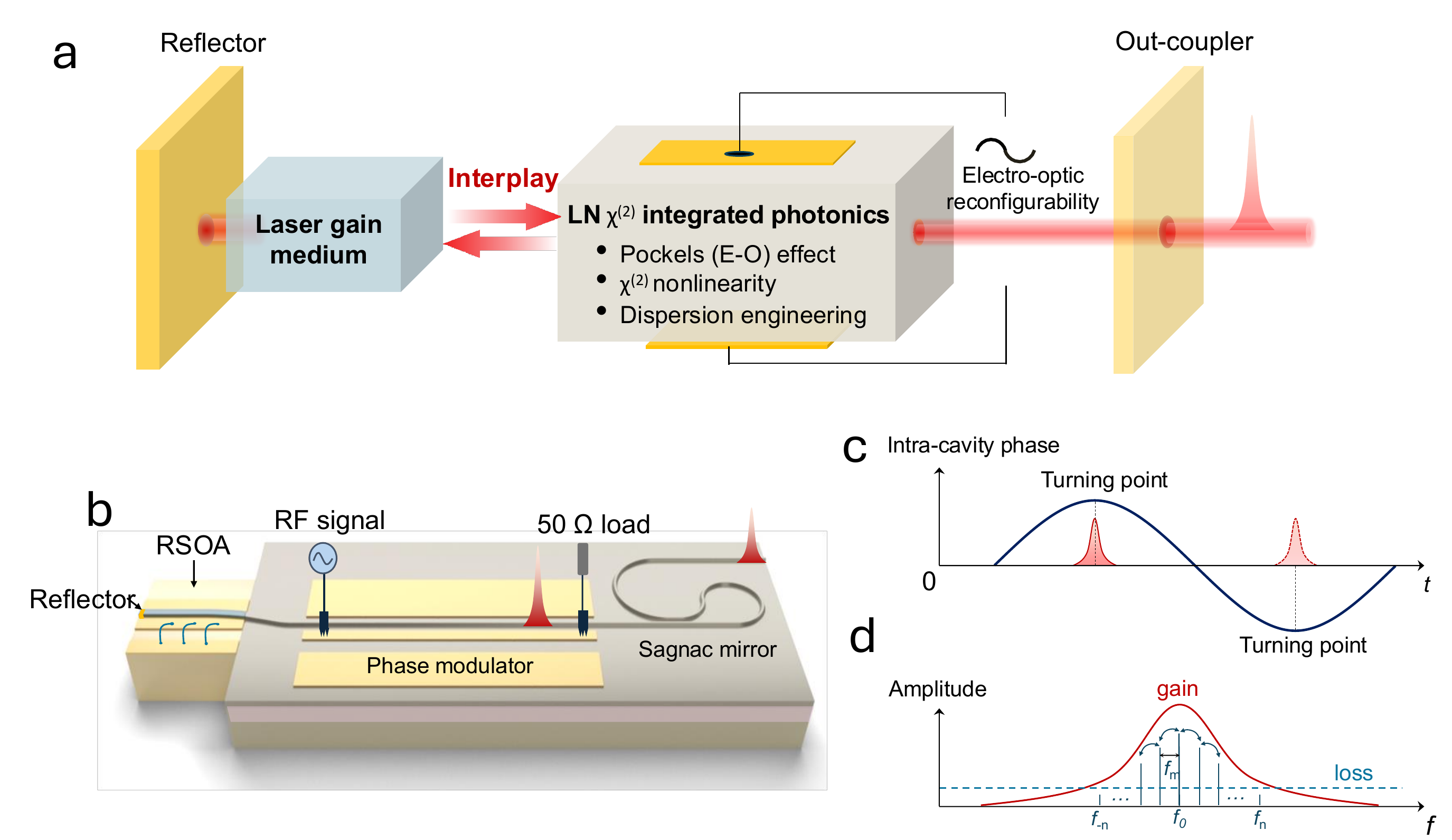}

\caption{MLL on LN integrated photonics. (a) Concept of integration of gain medium with LN quadratic integrated nonlinear photonics for developing chip-scale MLLs. (b) Schematic of the integrated actively MLL on LN. (c) Illustration of mode-locking in the time domain. (d) Illustration of active mode-locking in the frequency domain. (b), (c) and (d) are adapted from Ref. \cite{guo2023ultrafast}.} \label{MLL} 
\end{figure*}
\vspace{-10 pt}

\subsubsection{Integrated LN ultrafast mode-locked lasers}

Mode-locked lasers (MLLs) are the most commonly used coherent light sources for producing extremely short optical pulses, typically ranging from picoseconds to femtoseconds. Their operation is based on phase-locking the longitudinal modes of the laser cavity so that their constructive interference generates a train of high-intensity periodic pulses in the time domain \cite{keller2021ultrafast}. As sketched in Fig. \ref{MLL}a, the synergistic integration of advanced gain media (e.g., III-V semiconductors or solid-state gain media) with LN $\chi^{(2)}$ integrated nonlinear photonics represents a promising pathway to realizing chip-scale actively or passively MLLs with potentially superior performance compared to existing approaches.
\vspace{1.5mm}

Beyond the aforementioned E–O frequency combs, the efficient (low-V$_\pi L$) Pockels effect in TFLN can be harnessed to realize on-chip active mode-locking elements, such as an intra-cavity E-O phase modulator (PM) or intensity modulator (IM). For instance, an actively MLL can be achieved through the hybrid integration of a III-V semiconductor gain element with a TFLN extended cavity incorporating an integrated broadband Sagnac mirror and an intra-cavity phase modulator \cite{guo2023ultrafast,ling2024electrically,zhang2026broadband}, as illustrated in Fig. \ref{MLL}. In this configuration, a Fabry-Perot laser cavity is formed between the rear reflector of the III-V gain chip and the Sagnac mirror on TFLN. When a sinusoidal RF driving signal is applied to the LN phase modulator, the Pockels effect periodically modulates the refractive index of the LN cavity. This is equivalent to periodically modulating the cavity length through a “moving end mirror”. When a circulating pulse (such as a noise spike) encounters this mirror while moving, it is reflected with a Doppler frequency shift \cite{siegman1986lasers}. After successive round trips, the accumulation of such Doppler shifts precludes a steady-state solution. However, when perfect synchronization is established between the E-O modulation and the cavity round-trip time, a circulating pulse strikes the mirror precisely at the ``turning points'' where the phase modulation reaches its extrema (Fig. \ref{MLL}c). In this case, the pulse only acquires a small quadratic phase modulation, or chirp, accompanied by spectral broadening from sideband generation. Consequently, a steady-state ultrashort pulse can be sustained in the cavity over many round-trips. 
\vspace{1.5mm}

In the spectral domain, the sidebands of each longitudinal mode generated by RF modulation overlap with adjacent longitudinal modes, enforcing all the longitudinal modes mutually injection-locked. While the quadratic phase modulation and side-band generation appear similar to those in E-O frequency combs described in Section \ref{section: E-O comb}, the underlying principle of active MLLs is fundamentally different. In actively MLLs, the phase-locked longitudinal modes will lase due to the presence of gain within the cavity, whereas in E-O comb sources, the comb is generated solely by redistributing energy from a single pump laser line \cite{hu2025integrated}. Consequently, unlike E-O combs discussed in Section \ref{section: E-O comb}, actively MLLs do not require a large RF modulation index to produce ultrashort pulses, and can therefore be realized in a much more compact form with low RF power consumption. Moreover, high pulse peak power can be achieved. Using this approach, peak powers beyond 0.5 W have been achieved at around 1060 nm with a repetition rate of $\sim$ 10 GHz \cite{guo2023ultrafast}.
\vspace{1.5mm}

As discussed above, the actively MLL with an intra-cavity PM tends to generate two pulses per round-trip. This is because pulses can be stably formed at either of the two phase modulation extrema and acquire chirps of opposite sign. Other dispersion mechanisms in the laser cavity, such as gain dispersion and cavity dispersion, may compensate the chirp of one pulse while further broadening the other, causing one pulse to dominate in peak power \cite{nagar1992pure}. By contrast, an actively MLL with an intra-cavity intensity modulator can enforce the generation of a single pulse per round trip with high peak power. Higher peak power can also be achieved by increasing cavity length using spiral waveguide geometries within the cavity \cite{qiu2025high,gao2025tightly}.
\vspace{1.5mm}

\begin{figure}[h]
\centering
\includegraphics[width=1\linewidth]{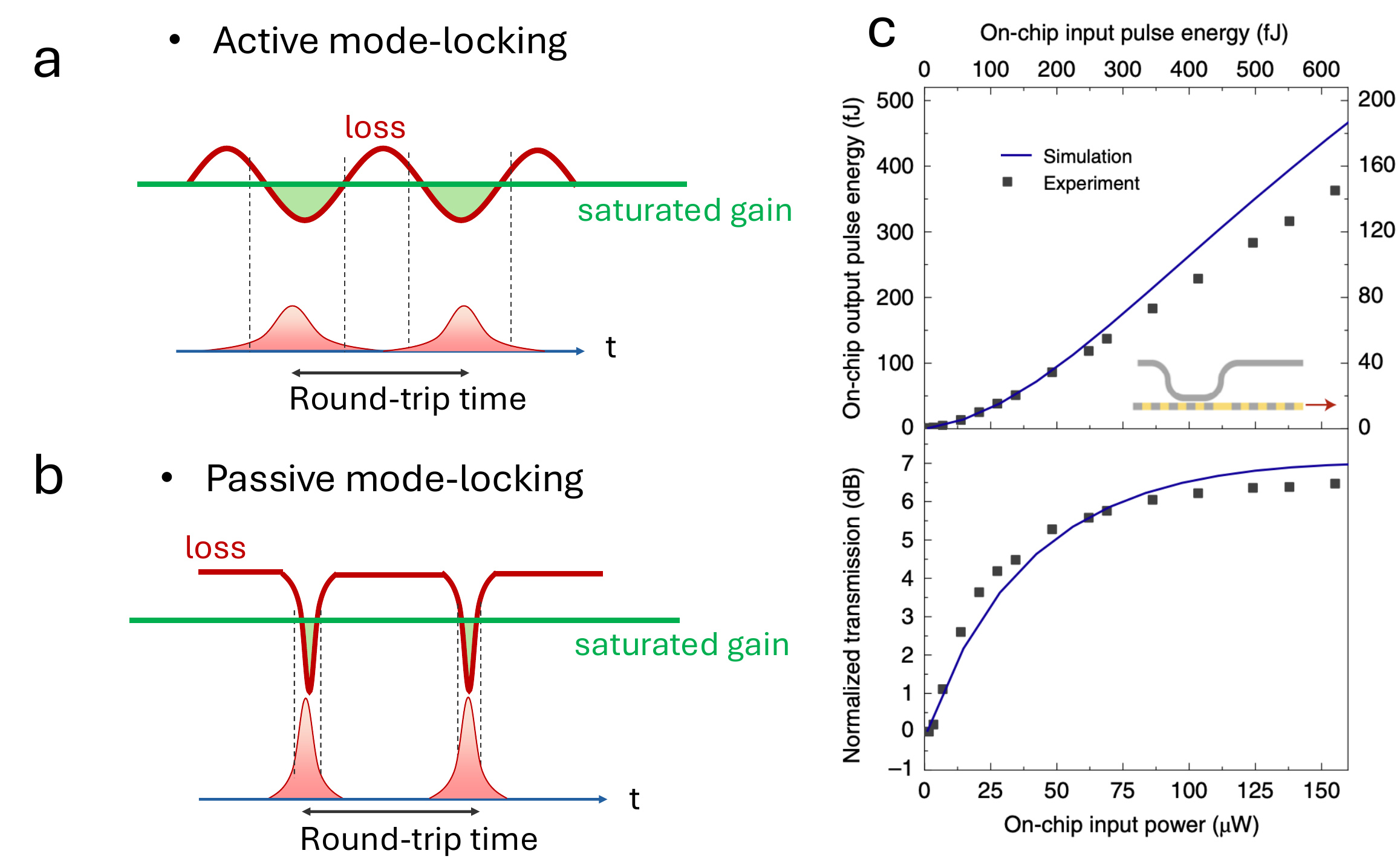}

\caption{Illustration of the net gain windows (shaded green area) for (a) active and (b) passive mode-locking. (c) Top: output pulse energy of the FH from the  $\chi^{(2)}$ integrated nonlinear splitter versus input average power/pulse energy. Bottom: normalized transmittance of the FH from the $\chi^{(2)}$ nonlinear splitter. Black symbols are the measured data, and blue solid curves are simulation results. Inset: schematic of the $\chi^{(2)}$ integrated nonlinear splitter. (c) is adapted from Ref. \cite{guo2022femtojoule}.} \label{MLL2} 
\end{figure}

While active mode-locking can provide high stability and low timing jitter through injection locking of adjacent axial modes, it generally yields relatively long pulses (more than a few picoseconds). Intuitively, this can be understood from the large net gain window inherent to active mode-locking (Fig. \ref{MLL2}a) \cite{haus2003theory,haus2002mode,keller2021ultrafast}, under the assumption that the laser gain recovery time is much longer than the pulse width and that the gain is fully saturated. To achieve even shorter pulses with higher peak intensity, passive mode-locking techniques—based on intra-cavity saturable absorbers \cite{wang2025passive} or artificial saturable absorbers—can be employed, since the effective net gain window in passive mode-locking can be much narrower, and can be even defined by the pulse width itself (Fig. \ref{MLL2}b). 
\vspace{1.5mm}

In this context, TFLN can also serve as an excellent integrated platform for passive mode-locking. Its energy-efficient and instantaneous $\chi^{(2)}$ nonlinearity and engineerable extinction ratio facilitate the generation of shorter pulses than those attainable with conventional semiconductor saturable absorbers. For instance, by minimizing the GVD and GVM and employing a nonlinear splitter geometry (Fig. \ref{MLL2}c inset) that controllably cascades the phase-matched SHG and the OPA sections in a PPLN nanophotonic waveguide \cite{guo2022femtojoule}, one can realize an ultrafast artificial saturable absorber (or nonlinear optical switch) with tens of femtosecond response time and femtojoule-level saturation energy (Fig. \ref{MLL2}c). Alternatively, the phase-mismatched cascaded $\chi^{(2)}$ process in a slightly phase-mismatched PPLN nanophotonic waveguide and the resulting efficient self-phase modulation of the fundamental pulses \cite{jankowski2020ultrabroadband,wise2002applications} can be exploited to enable integrated artificial saturable absorbers \cite{singh2024silicon,singh2020towards}. 

\subsubsection{$\chi^{(2)}$ supercontinuum generation and two-color $\chi^{(2)}$ soliton pulse compression}\label{Section:SCG}

So far, our discussion has focused on generating ultrashort light pulses from a CW pump light or gain medium. Now, we turn to the case of a pulsed pump. Supercontinuum generation (SCG) is a nonlinear optical process that produces an extremely broad and continuous spectrum of light with phase coherence between each frequency. Recently, there has been growing interest in exploiting $\chi^{(2)}$ interactions in dispersion-engineered PPLN nanophotonic waveguides as an alternative pathway for spectral broadening, owing to their unique advantages in pump energy efficiency compared to $\chi^{(3)}$ nonlinearities and their inherent capability for octave-spanning bandwidths. 
\vspace{1.5mm}

\textbf{$\chi^{(2)}$-based SCG:} Two mechanisms—though sometimes intertwined—can contribute to $\chi^{(2)}$-based SCG in a dispersion-engineered (low GVD and GVM) PPLN nanophotonic waveguide. The first mechanism is an effective $\chi^{(3)}$ nonlinearity arising from cascaded $\chi^{(2)}$ interactions in an undepleted or weakly-depleted regime.  Specifically, when phase-mismatch ($\Delta k$) exists, the light pulse undergoes repeated up- and down-conversion between the fundamental-harmonic (FH) and second-harmonic (SH) fields along the propagation direction. The back conversion of the SH to FH will induce an intensity-dependent phase shift (i.e., self-phase modulation (SPM)) of the FH, giving rise to the spectral broadening of both harmonics \cite{liu1999high}. In this case, the coupled-wave equations describing the SHG process can be reduced to the standard nonlinear Schrodinger equation with an effective $\chi^{(3)}$. The amplitude of the effective $\chi^{(3)}$ is determined by $|\Delta k|$ and the sign of the effective $\chi^{(3)}$ is determined by the sign of $\Delta k$ \cite{liu1999high,wise2002applications}. Notably, the effective $\chi^{(3)}$ nonlinearity can be two orders of magnitude larger than the intrinsic $\chi^{(3)}$ nonlinearity of LN \cite{jankowski2020ultrabroadband}. This enhanced nonlinearity greatly reduces the required pump pulse energy for SCG. Furthermore, at even higher pump powers, the interaction of both second-order and third-order nonlinearities can lead to a cascade of mixing processes that generate up to fifth-order harmonics \cite{hamrouni2024picojoule}. All the harmonics will undergo spectral broadening and finally emerge to form a multi-octave, coherent supercontinuum. 
\vspace{1.5mm}

\begin{figure}[h]
\centering
\includegraphics[width=1\linewidth]{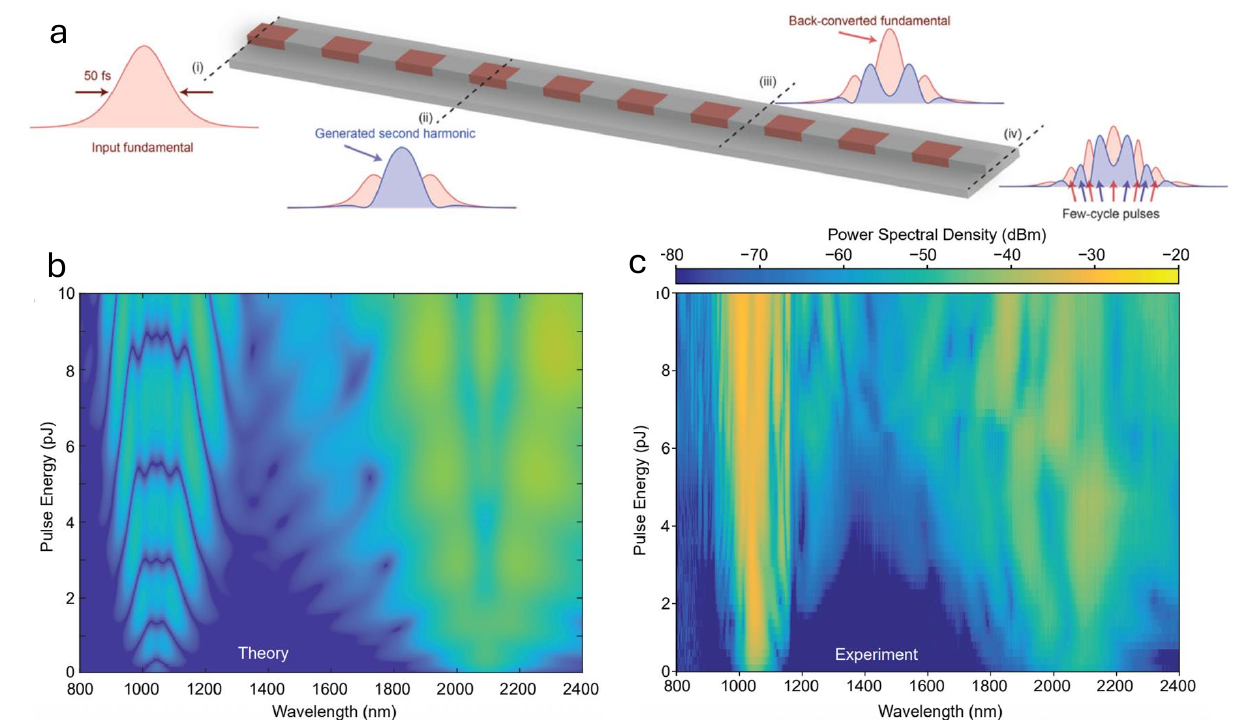}
\caption{Supercontinuum generation based on saturated $\chi^{(2)}$ interaction in PPLN nanophotonic waveguide. (a) Evolution of the fundamental and second harmonic pulses inside the PPLN nanophotonic waveguide.  (b) Simulated and (c) measured output pulse spectra under different pump pulse energies. (a) to (c) are adapted from Ref.\cite{jankowski2023supercontinuum}.} \label{Fig:SCG} 
\end{figure}
\vspace{-15 pt}

Another important mechanism is saturated $\chi^{(2)}$ interactions accompanied by strong pump depletion, which is dominant when the phase mismatch $\Delta k$ is close to zero \cite{jankowski2023supercontinuum,guo2022femtojoule}. In this regime, due to the low GVM, the SH and FH pulses are strongly temporally overalapped so that the generated SH pulse in a PPLN nanophotonic waveguide can become sufficiently intense to significantly deplete the FH pulse. As illustrated in Fig. \ref{Fig:SCG}a, this depletion may produce a pronounced temporal dip in the FH pulse—as regions of higher intensity undergo stronger SHG conversion. With increasing propagation length, continuous energy exchange between the FH and SH fields gives rise to rapid femtosecond-scale intensity oscillations, or even few-cycle pulse structures, within both the FH and SH pulses \cite{jankowski2023supercontinuum}, as depicted in Fig. \ref{Fig:SCG}a. In the frequency domain, the spectral broadening stems from these ultrafast temporal intensity variations and scales approximately with propagation distance, as the number of few-cycle features grows along the waveguide. Moreover, interference among these few-cycle pulses produces finely modulated fringes in the output spectra, as shown in Fig. \ref{Fig:SCG}b and c. The saturated $\chi^{(2)}$ interaction mechanism exhibits quadratic scaling of the required pump energy with waveguide length (linear scaling for $\chi^{(3)}$-based SCG), which allows the generation of octave-spanning SCG with remarkably low pulse energies while maintaining high coherence across an octave-spanning bandwidth.
\vspace{1.5mm}

Compared with octave-spanning SCG based on $\chi^{(3)}$ nonlinearities, the aforementioned $\chi^{(2)}$-SCG in dispersion-engineered PPLN nanophotonic waveguides not only easily enables coherent octave-spanning spectra, but also strong spectral overlap between the FH and SH combs. This overlapping harmonic comb leads to pronounced interference and beat notes that remain phase-locked across a broad bandwidth. Such a feature permits direct carrier–envelope offset (CEO) frequency ($f_\mathrm{ceo}$) detection immediately after the waveguide using a fast photodetector \cite{jankowski2020ultrabroadband,wu2024visible}, as a significant portion of the optical bandwidth can contribute to the photocurrent. This can greatly simplify $2f$ interferometry for $f_\mathrm{ceo}$ detection, thereby advancing the development of self-referenced frequency combs.
\vspace{1.3mm}

Another unique feature of $\chi^{(2)}$-based SCG is its ability to greatly expand the spectral coverage by simultaneously accessing multiple $\chi^{(2)}$ nonlinear processes within the dispersion-engineered PPLN nanophotonic waveguide, such as SHG, SFG, and DFG. For instance, by employing a single poling period together with a flat dispersion profile with two widely separated zero-dispersion wavelengths, an extremely large phase-matching bandwidth can be achieved. This allows for broadband DFG-assisted $\chi^{(2)}$-based SCG, which yields a remarkable 3.8-octave spectral coverage across the entire transparency window of lithium niobate, from 400 nm to 5 $\mu$m \cite{zhou2025quadratic}. Moreover, a recent work has demonstrated mid-IR SCG from 3.2 to 4.8 $\mu$m using a ``dual-stage architecture'' on LN that combines $\chi^{(3)}$ nonlinear spectral broadening with a PPLN section for efficient broadband intra-pulse DFG \cite{ludwig2025mid}.
\vspace{1.5mm}

\begin{figure}[h]
\centering
\includegraphics[width=1\linewidth]{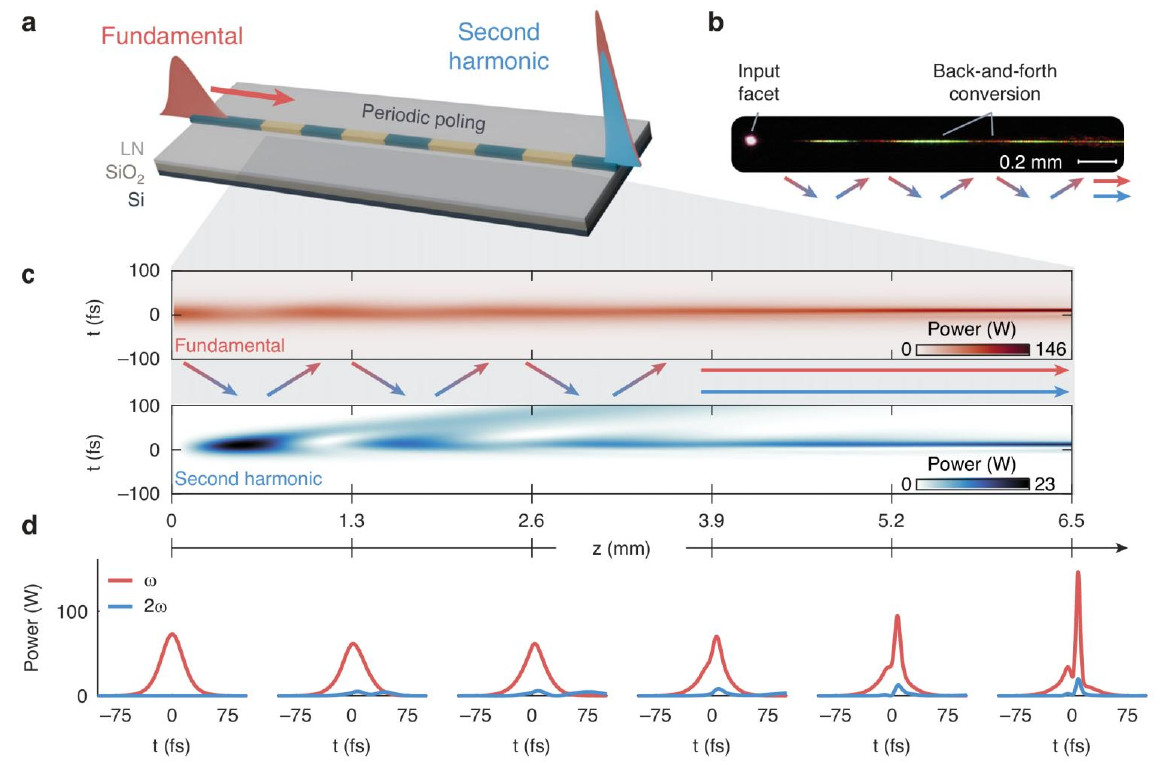}
\caption{Two-color soliton pulse compression on phase-mismatched PPLN nanophotonic waveguide. (a) Schematic of the two-color soliton pulse compression. The input fundamental pulse is compressed and accompanied by a co-propagating compressed SH pulse. (b) Microscope image of the the PPLN nanophotonic waveguide, showing the back-and-forth conversion between the FH and SH pulses. (c) Simulated evolution of the FH (top) and SH (bottom) pulses along the waveguide. (d) Temporal profiles of the FH and SH pulses at the labeled locations in (c). (a)-(d) are adapted from Ref. \cite{gray2025two}.} \label{soliton} 
\end{figure}

\textbf{Two-color $\chi^{(2)}$ soliton pulse compression:} The octave-spanning $\chi^{(2)}$-SCG holds great potential for generating few- or even single-cycle optical pulses. However, achieving the desired temporal pulse shape typically requires additional phase compensation. As an alternative approach, generation of clean, few-cycle ultrashort pulses can be achieved using two-color $\chi^{(2)}$ soliton pulse compression in dispersion-engineered PPLN nanophotonic waveguides \cite{gray2025two}, where a temporally broad FH pulse is compressed to a much shorter duration accompanied by the generation of a short co-propagating SH pulse, as illustrated by Fig. \ref{soliton}a. Specifically, in the regime of low-GVM, low-GVD, and near phase matching ($\Delta k \sim 0$), FH and strongly on-chip generated SH pulses remain strongly temporally overlapped, enabling continuous energy exchange (back-and-forth conversion) during propagation (Fig.\ref{soliton}b). This interaction induces saturated $\chi^{(2)}$ spectral broadening, similar to that described in Section~\ref{Section:SCG}. Two-color $\chi^{(2)}$ solitons and efficient pulse compression occurs when this nonlinear spectral broadening effect is just balanced by the temporal broadening effect by the linear GVD. Notably, the significantly reduced temporal walk-off in dispersion-engineered PPLN waveguides allows this scheme to overcome the pulse-compression limits of conventional $\chi^{(2)}$ soliton compression in the limit of large phase-mismatch \cite{bache2007scaling,liu1999high}.
\vspace{1.5mm}

As shown in Fig. \ref{soliton}c and d, the two-color $\chi^{(2)}$ soliton compression dynamics arise from periodic pulse broadening and compression associated with high-order soliton evolution. The achievable compression factor is governed by the soliton number, which depends on the input pulse peak power and duration: larger soliton numbers yield stronger compression but reduced pulse quality \cite{bache2007scaling}. With optimized waveguide length and pump power, Gray et al. experimentally demonstrated compression of a 48-fs pulse to 13 fs (sub-two-cycle at 2090 nm), while the SH pulse reached 16 fs~\cite{gray2025two}. By controlling the relative phase between the FH and SH pulses—e.g., via carrier-envelope-phase tuning using an integrated E–O modulator—single-cycle pulse synthesis becomes feasible.

\subsubsection{Synchronously pumped $\chi^{(2)}$ OPO}\label{Section:sync pumping}

In addition to functioning as wavelength-tunable CW light sources as described in Section \ref{section: CW OPO}, integrated OPOs on TFLN can also enable ultrashort pulse or soliton formation purely based on $\chi^{(2)}$ nonlinearity. Compared to Kerr soliton microcombs, $\chi^{(2)}$ OPOs can allow pulse and soliton formation at substantially lower pump powers or with reduced cavity $Q$-factors. Furthermore, through parametric down-conversion, they facilitate pulse generation in spectral regions where Kerr soliton combs are difficult or even inaccessible—especially in the mid-infrared.
\vspace{1.5mm}

\begin{figure*}[ht]
\centering
\includegraphics[width=1\linewidth]{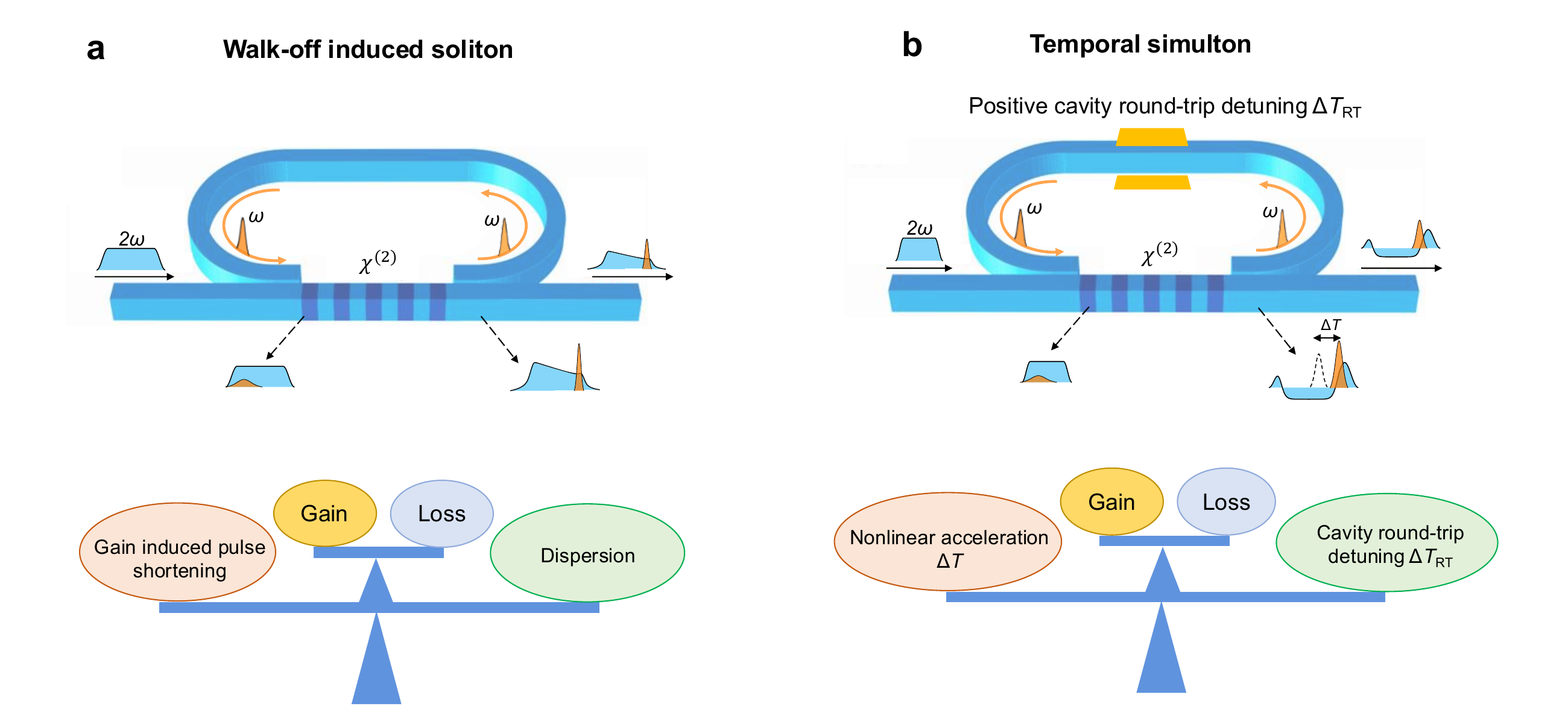}
\caption{\textbf{$\chi^{(2)}$ temporal solitons in non-stationary SPDOPOs}. (a) Upper panel: schematic of walk-off-induced soliton formation in SPDOPO with a large GVM. Owing to the large GVM between the signal and the pump, substantial pulse compression occurs due to the non-stationary OPA process as the signal walks off through the pump. Lower panel: double balance of temporal broadening and energy for walk-off induced solitons. (b) Upper panel: schematic of temporal simulton formation in SPDOPO with a small GVM. After OPA, the signal ($\omega$) acquires a nonlinear shift in group delay $\Delta T$, which is compensated by the timing mismatch $\Delta T_\mathrm{RT}$ due to positive cavity round-trip detuning. Lower panel: double balance of timing and energy for simulton formation. } \label{OPO soliton} 
\end{figure*}

Since $\chi^{(2)}$ nonlinearity is more pronounced with strong instantaneous field intensities, and large field intensities are most readily achieved in the pulsed operation, there has been growing interest in the synchronously pumped OPO \cite{van1995synchronously, leindecker2011broadband}, in which the pump is a train of ultrashort pulses synchronized or almost synchronized to the round-trip time of the OPO cavity. One promising system is the synchronously pumped degenerate OPO (SPDOPO), in which a $\chi^{(2)}$ resonator pulse-pumped at 2$\omega$ generates a resonant half-harmonic (signal/idler) at $\omega$. The advantage of operating at the degeneracy is threefold. First, due to the phase-sensitivity of the OPO at the degeneracy, the signal comb is phase-locked to the pump comb \cite{wong2010self}. Second, the large optical parametric gain bandwidth around the degeneracy facilitates the generation of shorter pulses or broader comb \cite{leindecker2011broadband}. Third, owing to the phase-sensitive parametric gain of the OPO at degeneracy and the presence of vacuum fluctuations, SPDOPOs can generate light pulses with binary (0 or $\pi$) phases, enabling their use as high-data-rate optical binary random bit generators {\cite{gray2025large,marandi2014network,mcmahon2016fully,inagaki2016coherent} and probabilistic bits \cite{roques2023biasing,horodynski2025stochastic,choi2024photonic}} for optical computing and sensing applications, as well as sources for squeezed vacuum  \cite{stokowski2023integrated,park2024single}. In the following, we discuss the different operating regimes of SPDOPOs and the distinct device behaviors and applications they enable. 
\vspace{1.2mm}

\textbf{Stationary regime:} The stationary regime of the SPDOPO corresponds to the condition where temporal walk-off between the pump pulse at $2\omega$ and the signal/idler pulse at $\omega$ is negligible. This ensures their temporal overlap throughout the interaction length. This occurs when $\text{GVM} \times L$ is smaller than $\tau_p$, where $L$ is the interaction length between the pump and signal pulses and $\tau_p$ is the pump pulse duration. In this regime, provided that the pump peak power is moderate (not significantly higher than the threshold) and the GVDs of both pump and signal pulses are not significant, the generated signal pulses from the SPDOPO would largely inherit the temporal width of the pump pulses \cite{hamerly2016reduced,roy2023visible}. However, when the pump pulse is ultrashort and its peak power exceeds the OPO threshold by a certain amount, strong back-and-forth conversion between the 
2$\omega$ and $\omega$ pulses within the PPLN region (saturated  $\chi^{(2)}$ interactions described in section \ref{Section:SCG}) induce substantial spectral broadening of the OPO output around both $\omega$ and 2$\omega$. However, the broadened OPO spectrum is incoherent due to the accumulation of the nonlinear phase. Very intriguingly, further increasing pump power not only leads to an even broader OPO spectrum, but also can cause the generated OPO to be coherent again \cite{sekine2025multi}. This is because the slight detuning between the pump repetition period and cavity round-trip time can also cause a phase accumulation to just compensate that from the nonlinear effect. Leveraging this mechanism, 2.6-octave SCG has been demonstrated with a remarkably low pump-pulse energy of 121 fJ \cite{sekine2025multi}.
\vspace{1.5mm}

\textbf{Walk-off induced temporal soliton regime:} In an SPDOPO with a large GVM between the pump and signal pulses, in combination with long pump pulses (e.g. several picoseconds long), the interaction between pump and signal in the $\chi^{(2)}$ crystal becomes \textit{non-stationary}, and the signal pulse can even ``walk through'' the pump pulse in the PPLN section. This process can lead to significant gain-induced signal pulse shortening effects due to the ``gain-clipping effect'' \cite{roy2022temporal,hamerly2016reduced}, as illustrated by Fig. \ref{OPO soliton}a upper panel. In the presence of dispersion and loss, a temporal soliton named walk-off induced dissipative solitons can be sustained in the OPO \cite{roy2022temporal}, in which the energy loss through dissipation is balanced by the parametric gain, while the temporal broadening due to the cavity GVD is counteracted by the gain-induced pulse shortening effect, as illustrated in Fig. \ref{OPO soliton}a lower panel \cite{roy2022temporal}. It has been experimentally shown that the formation of walk-off induced soliton allows for a pump-to-signal pulse-compression factor of $\sim$42, which compressed an 13.2-ps pulse at 775 nm to 316 fs at 1550 nm with a high conversion efficiency of $\sim$10$\%$ \cite{roy2022temporal}. The pulse compression factor can be further enhanced by flattening the dispersion profile (i.e., minimizing the wavelength dependence of the effective refractive index) and increasing the GVM or gain section length.
\vspace{1.5mm}

\textbf{Temporal simulton regime:} Although walk-off–induced temporal solitons enable significant pulse compression, achieving higher efficiency and peak power is challenging because their formation requires negligible nonlinear phase shift, i.e., operation only slightly above threshold with minimal pump depletion. As the pump power further increases, the stable signal pulse may broaden or break up due to the widening of the temporal gain window. To overcome these limitations, one can exploit another type of soliton in non-stationary OPOs, known as a \textit{temporal simulton} \cite{jankowski2018temporal,hamerly2016reduced}, which is characterized by the simultaneous and phase-locked formation of a bright soliton in the signal at $\omega$ and a dark soliton in the pump at $2\omega$. A temporal simulton emerges from two coupled balances: (i) an energy balance between the parametric gain and loss, and (ii) a timing balance between the nonlinear signal pulse acceleration (characterized by the temporal advancement $\Delta T$) induced by pump depletion and the cavity round-trip detuning ($\Delta T_\mathrm{RT}$). This unique mechanism allows the signal pulse to extract more energy from the pump (efficiency can exceed 50$\%$), leading to strong pump depletion and the generation of high-peak-power signal pulses without significant pulse distortion. The formation of stable temporal simulton with shorter pulse duration necessitates the minimization of GVD and a nonzero third-order dispersion (TOD). The TOD contributes to a shift in the centroid of the simulton and works with time mismatch to compensate for the nonlinear phase shift. 
\vspace{1.5mm}

The non-stationary operating regimes of the SPDOPO discussed above have important implications for developing highly efficient, high-peak-power (even up to the watt level) ultrashort pulse sources, even when starting from a long, quasi-CW, or even CW pump \cite{khurgin2008passive}. Intuitively, this can be understood from the fact that in these non-stationary SPDOPOs, the signal can extract a large amount of energy from the pump by temporally walking off from the pump pulse. In this context, the GVM effectively acts as a pump energy storage mechanism—analogous to the excited-state lifetime in a mode-locked laser gain medium.

\section{Outlook}
Recent rapid progress in coherent light sources on engineered TFLN has not only deepened our understanding of nonlinear and electro-optic light–matter interactions, but also opened numerous opportunities in sensing, metrology, quantum and classical computing, and information processing. Here, we outline several promising directions to inspire future research endeavors.
\vspace{1.5mm}

\textbf{Accessing UV-visible spectral region.} As elaborated in Section \ref{section: Ultrashort pulsed light sources}, integrated photonic circuits operating at visible and UV wavelengths are crucial for emerging applications such as portable optical and atomic clocks for positioning, navigation, and timing, quantum sensing, and underwater communications.  Although existing low-loss photonic platforms in the UV-visible range—such as germano-silicate\cite{chen2026towards}, silicon nitride \cite{wang2016frequency, morin2021cmos,chauhan2022ultra,chul2020chip,porcel2019silicon,sorace2019versatile,siddharth2022near}, alumina \cite{he2023ultra,west2019low,du2024high,sorace2019versatile,franken2023hybrid}, and HfO$_2$ \cite{jaramillo2025hfo2}—are effective for routing light, they generally lack the ability to actively modulate, amplify, or generate light in this spectral region. Recently, it has been demonstrated that TFLN and its closely related platform, thin-film lithium tantalate (TFLT), can exhibit low optical losses in the visible band even down to 6 dB/m \cite{desiatov2019ultra,yu2021ultralow}. Developing visible $\chi^{(2)}$ parametric light sources \cite{park2022high,sayem2021efficient}, optical amplifiers or supercontinuum generators on TFLN or TFLT to simultaneously access multiple wavelengths that are difficult to reach with conventional semiconductor lasers or frequency doublers would enable a new generation of on-chip UV-visible photonic technologies. In this context, more advanced dispersion engineering capabilities such as inverse design \cite{yang2023inverse,vercruysse2019dispersion} would greatly extend the spectrum coverage of $\chi^{(2)}$ parametric light sources. However, the higher energy of UV or visible photons leads to increased photo-refractive effect, which can cause potential problems such as optical  \cite{bryan1984increased}, DC bias drift of E-O amplitude modulators \cite{ren2025photorefractive}, resonance frequency drift \cite{jiang2017fast,he2019self,xu2021mitigating,ren2025photorefractive}, as well as impact of the phase matching condition that can distort the nonlinear transfer function and reduce the nonlinear frequency conversion efficiency \cite{wang2025impact}. Therefore, a deeper understanding of the undesired photo-refractive effects in TFLN and TFLT and effective strategies for their mitigation \cite{wang2025impact,xu2021mitigating,ren2025photorefractive,cai2025stable,chen2006wavelength,wang2025impact} will be critical to advancing the device performance and reliability.
\vspace{1.5mm}

\textbf{Toward shorter pulse duration and higher pulse peak power.} Recently, it has been established that many nonlinear and ultrafast optical functionalities can be realized on LN nanophotonic platform, including SCG \cite{jankowski2023supercontinuum,jankowski2020ultrabroadband,wu2024visible,hamrouni2024picojoule,sekine2025multi}, OPA \cite{chen2025high,ledezma2022intense,jankowski2020ultrabroadband,dean2025low}, OPO \cite{lu2019periodically,ledezma2023octave,roy2023visible,sekine2025multi,kellner2025low,yang2025degeneracy}, pulse compression \cite{roy2022temporal,gray2025two}, all-optical switching \cite{guo2022femtojoule}, all-optical nonlinear activation function \cite{li2023all}, and quantum
state generation \cite{nehra2022few,park2024single,stokowski2023integrated,williams2025ultrafast,shi2025squeezed, zhao2020high, jin2014chip, harper2024highly, finco2024time}. Therefore, developing highly coherent, high-peak-power, ultrashort pulsed light sources integrated on nanophotonic LN chips can enable integrated ultrafast photonic systems with unprecedented performance and functionality. As described in Section \ref{section: Ultrashort pulsed light sources}, ultrafast MLLs integrated on TFLN \cite{guo2023ultrafast,wang2025passive} represents a promising pathway toward this attractive goal. To attain even higher peak power and shorter output pulses, it is advantageous to integrate rare-earth ions such as Er$^{3+}$ \cite{liu2024fully,liu2022photonic,zhou2021chip,luo2021chip,chen2021efficient,wang2020incorporation,qiu2025high}, Yb$^{+}$ \cite{pak2020ytterbium,zhang2022chip,zhang2023chip,luo2022integrated}, or Ti:Sapphire  \cite{wang2023photonic,wang2024heterogeneous,yang2024titanium} gain media onto the TFLN platform. Compared with semiconductor gain media, which typically exhibit very short excited-state lifetimes on the order of nanoseconds \cite{coldren2012diode} and fast gain recovery, these solid-state gain media possess much longer (micro-or milli-seconds) excited-state lifetimes, allowing the laser pulses to extract more energy from the pump per round trip and achieving higher peak power and efficiency. Solid-state gain media are also free from the undesired transient carrier dynamics in semiconductors, which often restrict pulse durations to above 200 fs \cite{delfyett1994femtosecond}. Exploiting the $\chi^{(2)}$ nonlinearities of TFLN for passive mode-locking and pulse shortening could further enable the generation of even shorter pulses while greatly simplifying system complexity. Moreover, as we elaborated in Section \ref{Section:sync pumping}, another promising approach involves leveraging the $\chi^{(2)}$ temporal solitons in non-stationary $\chi^{(2)}$ SPDOPO. By introducing sufficiently large temporal walk-off between pump and signal \cite{khurgin2008passive}, high-peak-power ultrashort temporal solitons can be sustained using quasi-CW or even CW pumps with modest power. Moreover, other types of $\chi^{(2)}$ solitons \cite{englebert2025topological,nie2025dissipative}, the interplay between E-O effect and $\chi^{(2)}$ parametric gain \cite{hamerly2024hybrid,stokowski2024integrated,sanchez2025quadratic}, cascaded $\chi^{(2)}$ nonlinearity \cite{tang2024broadband} can also be explored for efficient ultrashort light pulse generation and spectral broadening.
\vspace{1.5mm}

\textbf{Emerging applications for quantum technologies.}
On-chip ultrafast light sources have fundamental applications for quantum photonic technologies. For example, to create pure photons, femtosecond to picosecond pump pulses are necessary to generate spectrally pure photons \cite{xin2022spectrally,kundu2025periodically}, which is a fundamental requirement for a scalable quantum hardware \cite{psiquantum2025manufacturable}. Typically, such pump pulses are generated by bulky off-chip lasers \cite{xin2022spectrally}. As the LN platform can be used as an efficient source of photon pairs \cite{xin2022spectrally,zhao2020high}, integration of ultra-fast pump lasers with a parametric photon pair source will enable a compact solution for photon pair generation for quantum computation and communication applications. Dispersion-engineered high-speed high-gain OPAs can also be used to amplify and measure single photons \cite{sendonaris2024ultrafast}. Current state-of-the-art quantum detection in photonic technology, either for communication or computation, utilizes superconducting nanowire single-photon detectors (SNSPDs) that require cryogenic cooling. SNSPDs are also inherently slow due to material constraints limiting the maximum detection bandwidth up to 1\,GHz \cite{resta2023gigahertz}. Dispersion-engineered OPAs on LN can offer high-speed detection, elevating the bandwidth bottleneck of SNSNDs, and most importantly, offer room-temperature detection. This may allow scalable room temperature solutions for quantum communication and computation applications not possible with other material platforms. 

\textbf{Emerging applications beyond light sources.} Another promising direction lies in harnessing the unique properties of TFLN or TFLT $\chi^{(2)}$ integrated light sources—such as ultra-wideband wavelength reconfigurability, phase-sensitive parametric amplification and oscillation, electro-optic modulation, $\chi^{(2)}$ nonlinear dynamics, as well as their mutual interplay—to unlock new capabilities in sensing, computing, and information processing. For example, the efficient E-O effect of TFLN or TFLT nanophotonic waveguides or resonators can enable the generation and detection of microwave \cite{hossein200614,zhu2025integrated,zhang2025integrated} and THz fields \cite{herter2025thin,herter2023terahertz,lampert2025photonics,cao2025integrated}, which may ultimately lead to a monolithic THz spectroscopic sensing system. The wide-band signal and idler wavelength tuning range of an integrated LN OPO and the accessibility to mid-IR wavelength can enable high-resolution on-chip mid-infrared spectroscopy \cite{hwang2023mid}. Moreover, the wide-band wavelength reconfigurability of an OPO can be leveraged to detect extremely small perturbations in the cavity without requiring a high-$Q$ optical resonator \cite{roy2021spectral,roy2021nondissipative,roy2023non}. The high OPO slope efficiency due to temporal simulton formation in SPDOPOs can substantially enhance the sensitivity of intra-cavity molecular sensing \cite{gray2024quadratic}. In addition, the probabilistic random-phase states of chip-integrated SPDOPOs can serve as high-rate probabilistic bits (p-bits) for probabilistic computing—an emerging computing architecture \cite{roques2023biasing,horodynski2025stochastic}.

\vspace*{0pt}

\begin{funding}
The authors acknowledge support from NSF Grant No.
2338798 (CAREER Award), the start-up grants from the CUNY Advanced Science Research Center and the CUNY Graduate Center, and
the PSC CUNY Research Award.
\end{funding}

\vspace*{-15pt}

\begin{authorcontributions}
All authors have accepted responsibility for the entire content of this manuscript and consented to its submission to the journal, reviewed all the results and approved the final\break version of the manuscript. MT and QG prepared the manuscript and figures, with contributions from all\break co-authors. QG drafted the outline and revised the manuscript.

\end{authorcontributions}

\vspace*{-14pt}

\begin{conflictofinterest}
The author states no conflicts of interest.
\end{conflictofinterest}

\vspace*{-15pt}

\begin{dataavailabilitystatement}
The data that support the plots within this paper and
other findings of this study are available from the corre-
sponding author upon reasonable request.
\end{dataavailabilitystatement}

\vspace*{-22pt}



\bibliographystyle{ieeetr}

\bibliography{references}


\clearpage

\end{document}